\pdfoutput=1
\documentclass[final,5p,times,twocolumn,authoryear,fleqn]{elsarticle}

\usepackage{placeins}
\usepackage{amssymb}
\usepackage{amsmath}
\usepackage[colorlinks=true,linkcolor=blue,citecolor=blue,urlcolor=blue]{hyperref}
\usepackage{booktabs}
\usepackage[table]{xcolor}
\usepackage{multirow}
\usepackage{graphicx}
\journal{New Astronomy}

\begin{document}

\begin{frontmatter}

\title{Maxwell--Bloch Modelling of Periodic 6.7~GHz Methanol Flares. I. Stable Quiescent Emission}

\author[first]{Toktam Rashidi}
\author[first]{Zhejian Liu}
\author[first]{Vahid Anari}
\author[second]{Yoshinori Yonekura}
\author[third]{Anna Bartkiewicz}
\author[third]{Paweł Wolak}
\author[first]{Fereshteh Rajabi}

\affiliation[first]{organization={Department of Physics and Astronomy, McMaster University},
addressline={1280 Main Street West}, city={Hamilton}, postcode={L8S 4L8}, state={Ontario}, country={Canada}}

\affiliation[second]{organization={Center for Astronomy, Ibaraki University},
addressline={2-1-1 Bunkyo}, city={Mito}, postcode={310-8512}, state={Ibaraki}, country={Japan}}

\affiliation[third]{
organization={Institute of Astronomy, Faculty of Physics, Astronomy and Informatics, Nicolaus Copernicus University in Toruń}, addressline={Grudziądzka 5}, city={Toruń}, postcode={87-100}, country={Poland}}

\begin{abstract}
We study periodic 6.7~GHz methanol flares in four maser-hosting regions with approximately stable quiescent emission. We test whether different velocity-component profiles within a source require different pump shapes or can arise from different local responses to a common symmetric pulse shape and width. Folded profiles span narrow to broad flares and include fast-rise/slow-decay, approximately symmetric, and slow-rise/fast-decay shapes. Selected well-sampled light curves with fast-rise/slow-decay or approximately symmetric profiles are modelled with the one-dimensional Maxwell--Bloch equations. For each velocity component, the measured flare period is imposed as the repetition period of the pump-pulse train. Within each source, the modelled components share the same symmetric pulse shape and width, while their effective medium parameters may differ and the individual pulse amplitudes may vary from cycle to cycle. Coherence diagnostics classify the fitted responses as maser-tending, superradiance-tending, or superradiance using sample-specific criteria. The flare morphology and radiation regime are therefore not determined by the imposed pump-pulse form alone, but also by the local inverted column density, the relaxation and dephasing timescales, and the dynamical state of the medium.
\end{abstract}


\end{frontmatter}

\section{Introduction}
\label{sec:introduction}

Periodic 6.7~GHz methanol maser light curves describe two complementary aspects of variability: the recurrence of the activity and the evolution of the emission within each cycle. The first is quantified by the period, whereas the second can be characterised by the flare width, amplitude, contrast, and relative rise and decay times. Long-term monitoring shows that these properties are not linked in a one-to-one manner: velocity components within a source can share a common period yet display profiles with different widths, strengths, and asymmetries \citep[e.g.,][]{Goedhart2004,Szymczak2018,Olech2019}. Flare morphology therefore provides information distinct from the recurrence period and should not be regarded as a direct proxy for the mechanism that establishes the periodicity.

The 6.7~GHz Class~II methanol transition is radiatively pumped by infrared emission from warm dust, which sustains the population inversion required for maser emission \citep{Sobolev1997,Cragg2005,Breen2013}. Proposed explanations for periodic variability generally invoke modulation of either the infrared pump or the background seed radiation. Possible origins include colliding-wind binaries, cyclic accretion, stellar pulsation, and spiral shocks in binary systems \citep{vanDerWalt2011,Araya2010,Inayoshi2013,Parfenov2014}. Such mechanisms may establish the recurrence timescale of the activity, but they do not by themselves determine the duration, shape, or radiation regime of the observed flare. Here, we do not attempt to identify the physical origin of the periodic modulation. Instead, we treat the periodic pump as an input and ask whether different flare profiles require distinct pump-pulse forms or can arise from different responses of the local medium to a common pump.

We address this question with the one-dimensional Maxwell--Bloch equations (MBEs), which trace the coupled evolution of the population inversion, macroscopic molecular coherence (or the polarisation of the medium), and radiation field. We represent the excitation as one symmetric pump pulse per flare cycle, with the repetition period fixed to the measured period of each component. Within each source, the modelled components share the same pulse shape and width, while their effective medium parameters and cycle-dependent pulse amplitudes may differ. This prescription fixes the pump cadence, shape, and width, while the flare duration and asymmetry emerge from the coupled matter--radiation dynamics.

The MBEs describe quasi-steady maser amplification and fast-transient superradiance as limiting behaviours of the same coupled light--matter dynamics \citep{Feld1980,Rajabi_2020_maser,Houde2024}. Near the quasi-steady maser limit, the molecular coherence adjusts rapidly to the evolving field and population inversion. Toward the fast-transient superradiance limit, the coherence develops its own dynamics before relaxation and dephasing can suppress the cooperative evolution, allowing the emitted flare to differ from the pump in both timing and profile. Intermediate responses connect these limits. Because the observed flare profile alone does not identify the realised regime, we assess the fitted solutions using diagnostics based directly on their coherence dynamics.

Previous applications of the Maxwell--Bloch framework have focused mainly on individual sources or events, including non-periodic 6.7~GHz methanol flares \citep{Rajabi2019,Fang2026}, multi-transition periodic variability in G9.62$+$0.20E \citep{Rajabi2023}, periodic flaring in G22.356$+$0.066 \citep{Rashidi2025}, and multi-periodic behaviour in G9.62$+$0.20E \citep{Rashidi2026}. The present work extends these studies to a comparative analysis of flare morphology and response dynamics across several sources and velocity components.

We analyse long-term monitoring data for four periodically flaring sources: G12.889$+$0.489, G14.230$-$0.509, G37.54$+$0.21, and G107.298$+$5.63. Representative folded profiles are used to measure the recurrent morphology, while selected multi-cycle intervals of the original light curves are fitted with the MBEs to retain cycle-to-cycle variations. Each modelled component is represented by a continuously integrated solution over its selected interval. For G37.54$+$0.21, a smoothly varying effective coherence-dephasing time $T_2$ is used within the continuous solution.

The fitted solutions span maser-tending, superradiance-tending, and superradiance behaviour. Despite the common pump prescription within each source, the fitted components develop different flare morphologies and coherence dynamics. The central result is therefore that the local matter--radiation response, rather than the imposed pump-pulse form alone, determines the emergent morphology and radiation regime.

This first paper considers components and modelling intervals with approximately stable quiescent emission. Components whose quiescent levels evolve substantially will be addressed in Paper~II.

Section~\ref{sec:data} describes the observations and source sample. Section~\ref{sec:profile_analysis} presents the period and morphology measurements. Section~\ref{sec:model} develops the Maxwell--Bloch model and coherence diagnostics, and Section~\ref{sec:results} presents the light-curve fits and diagnostic results. The implications and limitations are discussed in Section~\ref{sec:discussion}, and the main conclusions are summarised in Section~\ref{sec:conclusions}. The appendices provide details of the period search, folded-profile analysis, and numerical implementation, together with the fitted parameters and remaining coherence-diagnostic plots.
\section{Data and source sample}
\label{sec:data}

\subsection{Monitoring observations}
\label{sec:observations}

Most of the 6.7~GHz methanol maser light curves analysed in this work were obtained from the Ibaraki 6.7~GHz Class~II Methanol Maser Monitoring Database (iMet)\footnote{\url{https://vlbi.sci.ibaraki.ac.jp/iMet/}}. The programme began in December 2012 with the Hitachi 32-m radio telescope at Ibaraki Observatory \citep{Sugiyama2015,Yonekura2016,Yonekura2024}. The initial cadence was approximately 9~d per source, while subsequent observing intervals ranged from approximately 1 to 50~d \citep{Yonekura2024}. The iMet data used in this study span MJD~56294.161--61071.093, corresponding to a maximum baseline of approximately 13~yr. The observations used 5-min integrations over an 8-MHz band centred on the 6.7~GHz methanol transition. The analysed spectra have an effective velocity resolution of $0.13~\mathrm{km\,s^{-1}}$, a typical $1\sigma$ rms noise of approximately 0.3~Jy, and a half-power beam width of approximately $4.6^{\prime}$ \citep{Sugiyama2015,Yonekura2016,Yonekura2024}.

For G12.889$+$0.489, the iMet light curves were supplemented with observations from the ongoing Toruń monitoring programme, initiated in 2009 with the Toruń 32-m radio telescope \citep{Szymczak2018}. The Toruń observations used here span MJD~57490--61252 (2016 April--2026 July) and were obtained in frequency-switching mode, typically every 5--10~d, with occasional gaps of 3--4~weeks. The spectra have a velocity-channel spacing of $0.044~\mathrm{km\,s^{-1}}$, a typical $3\sigma$ sensitivity of $0.8$--$1.2$~Jy, and a flux-density calibration uncertainty below 10 per cent.

The iMet and Toruń measurements of G12.889$+$0.489 were combined for the period search, folded-profile measurements, and Maxwell--Bloch modelling. No relative flux-density rescaling was applied: at epochs observed by both telescopes, the matched component flux densities agreed within approximately 3--15 per cent across the four compared components and within 3--5 per cent for the two components included in the Maxwell--Bloch analysis.

The four-source sample was chosen to include components with approximately stable quiescent emission and to span a range of periods and flare morphologies.

\subsection{Selected sources}
\label{sec:sources}

\noindent G12.889$+$0.489 is associated with the high-mass protostellar object IRAS~18089$-$1732 and lies at a trigonometric parallax distance of $2.34^{+0.13}_{-0.11}$~kpc \citep{Xu2011}. Periodic 6.7~GHz methanol flaring with a period of $29.5\pm0.1$~d was previously reported \citep{Goedhart2009}.

\noindent G14.230$-$0.509 (referred to as G014.23$-$00.50 by \citealt{Sugiyama2017}) lies in the infrared dark cloud G14.225$-$00.506 within the M17 Southwest Extension (M17 SWex). We adopt the $1.98^{+0.14}_{-0.12}$~kpc parallax distance measured toward the M17 region \citep{Xu2011,Sugiyama2017}. Its periodic variability is strongly velocity selective: the component near $v_{\rm lsr}=25.30~\mathrm{km\,s^{-1}}$ flares with a period of 23.89~d, whereas the other spectral components do not exhibit the same periodic pattern \citep{Sugiyama2017}.

\noindent G37.54$+$0.21 (IRAS~18566$+$0408; referred to as G37.55$+$0.20 by \citealt{Araya2010}) is a massive protostellar source at a kinematic distance of 6.7~kpc \citep{Araya2010}. Quasi-periodic 4.8~GHz formaldehyde flares recur approximately every 237~d, with separations between successive flares ranging from 234 to 248~d. Correlated variability was also reported in the 6.7~GHz methanol component near $v_{\rm lsr}=87.8~\mathrm{km\,s^{-1}}$ \citep{Araya2010}.

\noindent G107.298$+$5.63 (IRAS~22198$+$6336) is an intermediate-mass young stellar object \citep{Sanchez-Monge2010} at a trigonometric parallax distance of $0.764^{+0.028}_{-0.026}$~kpc \citep{Hirota2008}. Its 6.7~GHz methanol emission flares with a period of approximately 34.4--34.6~d \citep{Fujisawa2014,Szymczak2016,Olech2020}. The emission becomes weak or undetectable between flares, with a reported peak-to-quiescent flux-density ratio greater than 30 \citep{Fujisawa2014}. Several 22~GHz water maser features also vary on the same 34.4-d cycle and generally alternate with the 6.7~GHz methanol flares \citep{Szymczak2016}.

\section{Period analysis and flare-profile measurements}
\label{sec:profile_analysis}

We determine component-specific periods and use them to construct representative folded profiles for empirical morphology measurements. Selected multi-cycle intervals of the original light curves are subsequently modelled with the Maxwell--Bloch equations in Section~\ref{sec:model}.

\subsection{Period determination and results}
\label{sec:periods}

We determined component-specific periods using Lomb--Scargle periodograms \citep{Lomb1976,Scargle1982}, removing slow baseline trends where necessary. For each component, the strongest peak was retained only if its LS power exceeded the component-specific empirical detection threshold. \ref{app:periods} describes the detrending, frequency grid, significance calculation, and period resolution, and presents the periodograms of the accepted G37.54$+$0.21 components.

Table~\ref{tab:periods} summarises the representative source-level periods and the velocity components with detected periodicity, together with published periods for comparison.

\begin{table*}[t]
\centering
\caption{Period-search results for the selected sources. The representative period is a descriptive mean of the accepted component-level Lomb--Scargle periods, weighted by their peak resolutions as described in \ref{app:periods}. The exact component periods adopted for folding are listed in Table~\ref{tab:metrics}, and published periods are included for comparison. For G37.54$+$0.21, the literature value is the characteristic recurrence interval of the quasi-periodic 4.8~GHz formaldehyde flares toward IRAS~18566$+$0408/G37.55$+$0.20, for which correlated variability was also reported in one 6.7~GHz methanol component.}
\label{tab:periods}

\footnotesize
\setlength{\tabcolsep}{5pt}
\renewcommand{\arraystretch}{1.15}

\begin{tabular}{@{}lccp{0.39\textwidth}@{}}
\toprule
Source & Representative period & Published period & Periodic velocity components \\
& (d) & (d) & ($\mathrm{km\,s^{-1}}$) \\
\midrule
G12.889$+$0.489 & 29.56 & 29.5$^{a}$ & 28.63, 32.72, 36.55, 39.27 \\
G14.230$-$0.509 & 23.89 & 23.89$^{b}$ & 25.28 \\
G37.54$+$0.21 & 247.01 & $\sim237^{c}$ & 79.62, 83.77, 84.55, 84.92, 86.38, 87.39, 87.82 \\
G107.298$+$5.63 & 34.40 & 34.40$^{d}$--34.60$^{e}$ & $-16.88$, $-16.07$, $-15.50$, $-10.89$, $-9.18$, $-8.51$, $-7.31$ \\
\bottomrule
\end{tabular}

\vspace{2pt}
\footnotesize
$^{a}$\citet{Goedhart2009}; $^{b}$\citet{Sugiyama2017}; $^{c}$\citet{Araya2010}; $^{d}$\citet{Fujisawa2014}; $^{e}$\citet{Szymczak2016}.
\end{table*}

We recover a 29.56~d cycle in G12.889$+$0.489 and a 23.89~d cycle in G14.230$-$0.509, consistent with the previously reported periods of 29.5~d and 23.89~d, respectively \citep{Goedhart2009,Sugiyama2017}. The 34.40~d period recovered for G107.298$+$5.63 matches the value reported by \citet{Fujisawa2014} and is close to the 34.60~d reported by \citet{Szymczak2016}.

In G37.54$+$0.21, seven methanol components show aligned periodogram peaks near 247~d, yielding a representative period of 247.01~d. The published timescale of approximately 237~d refers to quasi-periodic 4.8~GHz formaldehyde flares in the same region, whose individual recurrence intervals range from 234 to 248~d \citep{Araya2010}. The methanol period recovered here lies near the upper end of that range.

The component-specific periods listed in Table~\ref{tab:metrics} are adopted for the folded-profile analysis below and, for the modelled components, set the repetition periods of the imposed pump pulses.

\subsection{Flare-profile measurements and morphology results}
\label{sec:metrics}

For each periodic component with a sufficiently stable quiescent level, we constructed a representative folded profile over the interval listed in Table~\ref{tab:metrics}. Shorter, component-dependent intervals were used where the baseline evolved appreciably. We retained the original, undetrended flux densities so that the fitted quiescent level and flare amplitude refer directly to the observed emission. To reduce the influence of uneven sampling, measurements in each phase bin were averaged first within each cycle and then across cycles.

Each folded profile was fitted with an empirical split Gaussian, with equal weight assigned to all populated phase bins. The fit yields the quiescent flux density $F_{\rm q}$, flare amplitude $A_{\rm fl}$ above that level, full width at half maximum $\Delta t_{1/2}$, and rise-to-fall ratio $R_{\rm r/f}$. The fractional width $\Delta t_{1/2}/p$ and contrast $C_{\rm fl}=A_{\rm fl}/F_{\rm q}$ measure the flare's duration relative to its period $p$ and strength relative to the quiescent emission. Values of $R_{\rm r/f}<1$, $=1$, and $>1$ indicate fast-rise/slow-decay, symmetric, and slow-rise/fast-decay profiles, respectively.

\ref{app:metrics} gives the profile-construction procedure, split-Gaussian function, fit weighting, and formal uncertainty propagation; the folded profiles and fits appear in Figures~\ref{fig_folded_g12_g14} and \ref{fig_folded_g37_g107}. The quoted uncertainties are conditional on the adopted period, phase binning, and MJD interval and exclude sensitivity to these analysis choices.

Table~\ref{tab:metrics} lists the morphology measurements. The $v_{\rm lsr}=28.63~\mathrm{km\,s^{-1}}$ component of G12.889$+$0.489 meets the period-search criterion but is excluded because it lacks a sufficiently stable quiescent interval. Daggers identify components selected for Maxwell--Bloch modelling of their original multi-cycle light curves in Section~\ref{sec:model}; the folded profiles are not inputs to those fits.

\begin{table*}[t]
\centering
\caption{Morphology measurements from representative folded profiles. Here $p$ is the folding period, $R_{\rm r/f}$ is the rise-to-fall ratio, $\Delta t_{1/2}$ is the full width at half maximum measured above the quiescent level $F_{\rm q}$, $A_{\rm fl}$ is the flare amplitude above $F_{\rm q}$, $\Delta t_{1/2}/p$ is the fractional width, and $C_{\rm fl}=A_{\rm fl}/F_{\rm q}$ is the contrast. The final column gives the MJD interval used to construct each folded profile. Daggers identify components whose original light curves are modelled with the Maxwell--Bloch equations over selected intervals. Lower and upper uncertainty bounds are obtained from the 16th and 84th percentiles of the joint parameter draws, with the folding period held fixed.}
\label{tab:metrics}

\footnotesize
\setlength{\tabcolsep}{3pt}
\renewcommand{\arraystretch}{1.7}

\begin{tabular*}{\textwidth}{@{\extracolsep{\fill}}llcccccccl@{}}
\toprule
Source & $v_{\rm lsr}$ & $p$ & $R_{\rm r/f}$ & $\Delta t_{1/2}$ & $A_{\rm fl}$ & $F_{\rm q}$ & $\Delta t_{1/2}/p$ & $C_{\rm fl}$ & MJD interval \\
& ($\mathrm{km\,s^{-1}}$) & (d) & & (d) & (Jy) & (Jy) & & & \\
\midrule

\multirow{3}{*}{G12.889$+$0.489}
& $32.72^{\dagger}$ & 29.55 & $1.07^{+0.22}_{-0.22}$ & $11.63^{+1.03}_{-0.83}$ & $3.23^{+0.17}_{-0.21}$ & $10.84^{+0.17}_{-0.16}$ & $0.394^{+0.035}_{-0.028}$ & $0.298^{+0.019}_{-0.023}$ & 59000--61071 \\
& 36.55 & 29.56 & $1.75^{+0.44}_{-0.40}$ & $25.40^{+8.01}_{-9.24}$ & $1.89^{+0.87}_{-0.85}$ & $1.76^{+0.92}_{-0.87}$ & $0.86^{+0.27}_{-0.31}$ & $1.07^{+2.01}_{-0.68}$ & 58500--59500 \\
& $39.27^{\dagger}$ & 29.56 & $1.09^{+0.19}_{-0.20}$ & $12.32^{+0.95}_{-0.96}$ & $16.70^{+0.74}_{-0.97}$ & $60.55^{+0.82}_{-0.69}$ & $0.417^{+0.032}_{-0.032}$ & $0.276^{+0.014}_{-0.019}$ & 58000--60500 \\

\midrule

G14.230$-$0.509
& $25.28^{\dagger}$ & 23.89 & $0.61^{+0.04}_{-0.03}$ & $3.32^{+0.05}_{-0.04}$ & $3.51^{+0.04}_{-0.05}$ & $0.38^{+0.01}_{-0.02}$ & $0.1391^{+0.0021}_{-0.0018}$ & $9.17^{+0.45}_{-0.35}$ & 57000--60000 \\

\midrule

\multirow{7}{*}{G37.54$+$0.21}
& 79.62 & 246.78 & $3.23^{+1.75}_{-1.12}$ & $64.22^{+7.76}_{-6.62}$ & $0.63^{+0.05}_{-0.06}$ & $1.39^{+0.02}_{-0.03}$ & $0.260^{+0.031}_{-0.027}$ & $0.452^{+0.038}_{-0.046}$ & 56294--60000 \\
& $83.77^{\dagger}$ & 246.78 & $0.30^{+0.03}_{-0.03}$ & $66.32^{+1.78}_{-1.60}$ & $5.87^{+0.11}_{-0.13}$ & $3.74^{+0.07}_{-0.06}$ & $0.2688^{+0.0072}_{-0.0065}$ & $1.567^{+0.044}_{-0.049}$ & 57500--61071 \\
& $84.55^{\dagger}$ & 247.42 & $0.43^{+0.05}_{-0.04}$ & $75.82^{+2.53}_{-2.25}$ & $2.01^{+0.04}_{-0.05}$ & $1.75^{+0.03}_{-0.03}$ & $0.3065^{+0.0102}_{-0.0091}$ & $1.144^{+0.035}_{-0.041}$ & 57500--61071 \\
& $84.92^{\dagger}$ & 247.42 & $0.47^{+0.05}_{-0.04}$ & $81.89^{+3.02}_{-2.66}$ & $3.46^{+0.08}_{-0.09}$ & $4.22^{+0.06}_{-0.06}$ & $0.331^{+0.012}_{-0.011}$ & $0.819^{+0.025}_{-0.030}$ & 56294--61071 \\
& 86.38 & 247.42 & $0.38^{+0.09}_{-0.06}$ & $85.97^{+6.06}_{-5.36}$ & $1.10^{+0.05}_{-0.05}$ & $2.30^{+0.04}_{-0.04}$ & $0.347^{+0.024}_{-0.022}$ & $0.477^{+0.025}_{-0.029}$ & 56294--60000 \\
& $87.39^{\dagger}$ & 246.78 & $0.46^{+0.08}_{-0.06}$ & $45.36^{+2.08}_{-1.78}$ & $1.76^{+0.05}_{-0.06}$ & $0.55^{+0.03}_{-0.02}$ & $0.1838^{+0.0084}_{-0.0072}$ & $3.18^{+0.19}_{-0.21}$ & 56294--60000 \\
& $87.82^{\dagger}$ & 246.78 & $0.46^{+0.05}_{-0.03}$ & $46.77^{+1.20}_{-1.03}$ & $3.05^{+0.05}_{-0.06}$ & $0.83^{+0.03}_{-0.02}$ & $0.1895^{+0.0048}_{-0.0042}$ & $3.69^{+0.13}_{-0.16}$ & 56294--59900 \\

\midrule

\multirow{7}{*}{G107.298$+$5.63}
& $-16.88$ & 34.39 & $1.38^{+0.07}_{-0.05}$ & $4.18^{+0.04}_{-0.05}$ & $2.97^{+0.03}_{-0.03}$ & $0.37^{+0.01}_{-0.01}$ & $0.1216^{+0.0012}_{-0.0016}$ & $7.94^{+0.23}_{-0.20}$ & 58000--60000 \\
& $-16.07$ & 34.40 & $1.62^{+0.14}_{-0.15}$ & $4.46^{+0.12}_{-0.10}$ & $4.73^{+0.09}_{-0.11}$ & $0.51^{+0.03}_{-0.02}$ & $0.1298^{+0.0035}_{-0.0030}$ & $9.32^{+0.44}_{-0.62}$ & 58500--60500 \\
& $-15.50^{\dagger}$ & 34.40 & $0.79^{+0.04}_{-0.05}$ & $5.38^{+0.09}_{-0.08}$ & $4.32^{+0.06}_{-0.06}$ & $0.41^{+0.02}_{-0.03}$ & $0.1563^{+0.0027}_{-0.0023}$ & $10.42^{+0.74}_{-0.50}$ & 58300--59300 \\
& $-10.89$ & 34.41 & $1.43^{+0.09}_{-0.10}$ & $6.17^{+0.12}_{-0.10}$ & $1.92^{+0.03}_{-0.03}$ & $0.47^{+0.01}_{-0.01}$ & $0.1794^{+0.0034}_{-0.0029}$ & $4.13^{+0.12}_{-0.14}$ & 58150--60000 \\
& $-9.18^{\dagger}$ & 34.40 & $1.02^{+0.02}_{-0.02}$ & $4.61^{+0.03}_{-0.02}$ & $26.14^{+0.09}_{-0.12}$ & $0.45^{+0.03}_{-0.04}$ & $0.1341^{+0.0007}_{-0.0006}$ & $57.6^{+4.9}_{-3.7}$ & 58000--60000 \\
& $-8.51$ & 34.40 & $1.09^{+0.06}_{-0.05}$ & $5.93^{+0.07}_{-0.08}$ & $8.91^{+0.10}_{-0.09}$ & $0.46^{+0.04}_{-0.03}$ & $0.1724^{+0.0021}_{-0.0023}$ & $19.5^{+1.6}_{-1.7}$ & 58000--60000 \\
& $-7.31^{\dagger}$ & 34.40 & $0.89^{+0.02}_{-0.02}$ & $5.08^{+0.03}_{-0.03}$ & $88.55^{+0.35}_{-0.45}$ & $0.38^{+0.15}_{-0.12}$ & $0.1477^{+0.0009}_{-0.0008}$ & $234^{+113}_{-67}$ & 58000--60000 \\

\bottomrule
\end{tabular*}
\end{table*}

The measured profiles span a broad range of fractional widths, contrasts, and asymmetries. We use these properties descriptively, without imposing discrete morphological classes.

The $32.72$ and $39.27~\mathrm{km\,s^{-1}}$ components of G12.889$+$0.489 show broad, approximately symmetric flares ($R_{\rm r/f}\simeq1$) superposed on substantial quiescent emission. Their widths $\Delta t_{1/2}$ correspond to 39--42 per cent of a cycle, and their peak flux densities are only 28--30 per cent above the quiescent level ($C_{\rm fl}\simeq0.28$--$0.30$; Table~\ref{tab:metrics}). Despite its comparable repetition period, G14.230$-$0.509 exhibits a much narrower, higher-contrast profile, with $\Delta t_{1/2}$ corresponding to approximately 14 per cent of a cycle. The flare rises faster than it decays ($R_{\rm r/f}=0.61^{+0.04}_{-0.03}$), and its peak flux density is approximately 920 per cent above the quiescent level ($C_{\rm fl}\simeq9.2$).

Like G14.230$-$0.509, G107.298$+$5.63 exhibits narrow, high-contrast flares ($\Delta t_{1/2}/p\simeq0.12$--$0.18$; $C_{\rm fl}\simeq4$--$234$), including the highest contrasts in our sample. The largest contrast remains sensitive to uncertainty in the fitted quiescent level. Across its velocity components, however, the profiles range from fast-rise/slow-decay ($R_{\rm r/f}=0.79^{+0.04}_{-0.05}$), through approximately symmetric ($R_{\rm r/f}\simeq1$), to slow-rise/fast-decay ($R_{\rm r/f}=1.62^{+0.14}_{-0.15}$). These components provide an example of morphological diversity within a single source, with both asymmetry directions represented among narrow, high-contrast flares sharing a common period.

G37.54$+$0.21 has the longest repetition period in our sample ($p\simeq247$~d) and the largest flare widths ($\Delta t_{1/2}\simeq45$--86~d). Its profiles have fractional widths $\Delta t_{1/2}/p\simeq0.18$--0.35 and contrasts $C_{\rm fl}\simeq0.45$--3.7. Compared with G107.298$+$5.63, they generally occupy a larger fraction of the cycle and have lower contrast. Six of the seven profiles analysed for this source have fast-rise/slow-decay shapes ($R_{\rm r/f}=0.30$--$0.47$). The $79.62~\mathrm{km\,s^{-1}}$ component is the sole exception, with a best-fitting slow-rise/fast-decay profile ($R_{\rm r/f}=3.23^{+1.75}_{-1.12}$), although its asymmetry is less precisely constrained. Its width is comparable to that of the fast-rising $83.77~\mathrm{km\,s^{-1}}$ component ($\Delta t_{1/2}\simeq64$ and $66$~d, respectively).

Together, these measurements show that components with essentially the same period and comparable widths $\Delta t_{1/2}$ can have markedly different asymmetries, while sources with comparable periods can differ in fractional width and contrast. In Section~\ref{sec:model}, we use Maxwell--Bloch modelling of selected component light curves to investigate how the response of the local medium shapes the emergent flares. The fits use a common symmetric pump shape and width within each source, allowing us to explore the radiative responses and effective physical conditions compatible with the observations.

\section{Maxwell--Bloch analysis}
\label{sec:model}

We model the velocity components marked with a dagger in column~2 of Table~\ref{tab:metrics}. The fits use selected multi-cycle intervals of the original, undetrended light curves, preserving the observed cycle-to-cycle variations in the fitted data.

\subsection{Response model and fitting prescription}
\label{sec:mbe_model}

We map the $5_1A^+$ and $6_0A^+$ rotational levels of the 6.7~GHz methanol transition onto the upper and lower states of an effective two-level system, respectively. The corresponding population densities are $n_{\rm u}$ and $n_{\rm l}$. The Maxwell--Bloch equations (MBEs) describe the coupled evolution of the population-inversion density $n=n_{\rm u}-n_{\rm l}$, macroscopic polarisation amplitudes $P^\pm$ representing molecular coherence, and radiation electric-field amplitudes $E^\pm$. We use the rotating-wave and slowly varying-envelope approximations, with the retarded-time coordinate $\tau=t-z/c$. \ref{app:MBE} gives the equations, initial conditions, and field conventions.

Each component is represented by an elongated cylindrical sample of length $L$ and cross-sectional area $A$. We adopt a unit Fresnel number, $F=A/(\lambda L)=1$, where $\lambda$ is the transition wavelength. For this geometry, a one-dimensional treatment of propagation along the symmetry axis $z$ provides a suitable approximation \citep{Gross_Haroche_1982,Benedict_1996}. We evaluate the emergent intensity at the output boundary $z=L$.

The medium is characterised by its initial inversion and the effective population-relaxation and coherence-dephasing times, $T_1$ and $T_2$. These account for non-coherent processes, including population redistribution by inelastic collisions and phase disruption by elastic collisions \citep{Rajabi_2020_maser,Houde2024}. The parameter $n_0$ sets the spatially uniform initial inversion at $\tau=0$ through $n(z,0)=n_0\cos\theta_0\simeq n_0$, where $\theta_0$ is a small initial Bloch angle representing the quantum fluctuation that seeds emission in a finite molecular ensemble; its definition is given in \ref{app:MBE}.

Quasi-steady maser amplification and fast-transient superradiance are two limiting radiation regimes of the MBEs, with intermediate responses between them \citep{Feld1980,Rajabi_2020_maser,Houde2024}. 

In the maser limit, the polarisation remains close to its instantaneous steady-state value for the local field and inversion. In the superradiance limit, collective polarisation and inversion dynamics develop on timescales short compared with $T_1$ and $T_2$, allowing a rapid release of stored excitation. The emitted flare can therefore differ from the pump pulse in timing, width, and asymmetry.

We define the superradiance timescale associated with the initial inverted column density $n_0L$ as
\begin{equation}
T_{\rm R,0}=\frac{8\pi\tau_{\rm sp}}{3\lambda^2 n_0L},
\label{eq:tr0}
\end{equation}
where $\tau_{\rm sp}$ is the spontaneous-emission lifetime of a single molecule in the transition. For a freshly inverted sample, a necessary timescale ordering for cooperative growth is
\begin{equation}
T_{\rm R,0}<T_2.
\label{eq:timescale_condition}
\end{equation}

In the periodically driven solutions, each pump pulse acts on the inversion, polarisation, and field inherited from earlier cycles. We therefore use $T_{\rm R,0}$ as a comparative scale set by the initial condition, rather than as a prediction of an individual flare's width or peak time relative to the pump. We assess the radiation regime using the coherence diagnostics in Section~\ref{sec:diagnostics} and \ref{app:diagnostics}.

The total inversion pump consists of a constant background and one symmetric pulse per cycle:
\begin{equation}
\Lambda_n(\tau)=\Lambda_{n,0}+\sum_m\frac{\Lambda_{1,m}}{\cosh^2[(\tau-\tau_0-mp)/T_{\rm p}]},
\label{eq:pump1}
\end{equation}
where $\Lambda_{n,0}=n_0\cos\theta_0/T_1\simeq n_0/T_1$. The pulse full width at half maximum is
\begin{equation}
\Delta t_{p,1/2}=2\operatorname{arcosh}(\sqrt{2})\,T_{\rm p}\simeq1.763\,T_{\rm p}.
\label{eq:pump_fwhm}
\end{equation}

Within each source, the fits share the pulse shape and width $\Delta t_{p,1/2}$, with $p$ fixed to each component's measured period. Medium parameters are fitted separately for each component, and the pulse amplitudes $\Lambda_{1,m}$ may vary between components and cycles. The fitted offset $\tau_0$ aligns the pulse train with the observations.

This pump describes the effective local excitation without identifying its astrophysical origin; the flare width and asymmetry emerge from the coupled dynamics.

For each component, a single multiplicative scale factor converts the emergent intensity at $z=L$ to the observed flux density and is held fixed over the fitted interval. This factor absorbs the emitting solid angle, beaming, and number of unresolved radiating systems. The intrinsic intensity relative to the maser saturation intensity is therefore not constrained, and the saturation state of the maser-tending solutions remains outside the scope of this study.

Each component is integrated continuously over its selected interval. Only the G37.54$+$0.21 fits use time-dependent dephasing, accommodating profile changes not reproduced with the constant medium parameters explored. We describe the evolution of $T_2$ across this multi-cycle interval using two smooth transitions between three plateau values.\footnote{The hyperbolic tangent provides a smooth interpolation between plateau values; its form is phenomenological.} These transitions are not repeated each cycle, so different flares can evolve under different dephasing conditions:
\begin{equation}
T_2(\tau)=T_2^{(0)}+\sum_{k=1}^{2}\frac{T_2^{(k)}-T_2^{(k-1)}}{2}\left[1+\tanh\!\left(\frac{\tau-\tau_k}{\delta_k}\right)\right].
\label{eq:t2_steps}
\end{equation}
Here $T_2^{(k)}$ are the three plateau values, the transition centres $\tau_k$ correspond to epochs near MJD~57890 and 58640, and $\delta_k=100$~d sets the smoothing timescale. The inversion, polarisation, and field evolve without resets, while $n_0$, $L$, $T_1$, the pump width, and the flux-density scale remain fixed within each fit.

We fit the model light curves to the observations by manually adjusting the parameters, guided by the unweighted sum of squared residuals over each selected interval. The resulting solutions are representative rather than unique. Since the light curves constrain $n_0L$ more directly than $n_0$ and $L$ separately, we report $n_0L$ and $T_{\rm R,0}$ as the principal comparative quantities in Table~\ref{tab:params}.

Strongly slow-rise/fast-decay profiles remain in the empirical sample but are excluded from the reported fits. Within the parameter searches performed, acceptable fits were not obtained with a single effective sample driven by symmetric pump pulses. These profiles may reflect overlapping contributions from unresolved emitting regions at the same line-of-sight velocity, with differences in their response timing and strength producing the observed asymmetry.

\subsection{Coherence diagnostics}
\label{sec:diagnostics}

To assess the radiation regime of each fitted response, we test the quasi-steady polarisation approximation underlying maser amplification. Setting $\partial_\tau P^+=0$ in Equation~\eqref{eq:MBE_P}, using the instantaneous field, inversion, and dephasing time of the full solution, defines the instantaneous maser-limit polarisation $P^+_{\rm M}$:
\begin{equation}
P^+_{\rm M}=P^+_0+\frac{id^2T_2(\tau)}{\hbar}\,nE^-,
\label{eq:PM}
\end{equation}
where $d$ is the transition dipole magnitude and $P^+_0$ is the fixed seed polarisation defined in \ref{app:MBE}. This reference can vary with time as the local conditions evolve. The departure of the polarisation from this reference satisfies
\begin{equation}
P^+-P^+_{\rm M}=-T_2(\tau)\partial_\tau P^+.
\label{eq:P_departure_identity}
\end{equation}
This relation remains valid when $T_2$ varies with time and links the departure from the reference to the polarisation dynamics neglected in the quasi-steady approximation. 

We quantify the relative departure along the sample as
\begin{equation}
\epsilon_P(\tau)=\frac{\displaystyle\int_0^L|P^+-P^+_{\rm M}|\,dz}{\displaystyle\int_0^L\left(|P^+|+|P^+_{\rm M}|\right)\,dz+LP_{\rm floor}},
\label{eq:epsP}
\end{equation}
where $P_{\rm floor}$ is a positive floor with units of polarisation. The term $LP_{\rm floor}$ suppresses misleadingly large values of $\epsilon_P$ when both polarisation amplitudes are negligible. The diagnostic satisfies $0\leq\epsilon_P\leq1$. For appreciable polarisation, values near zero indicate a quasi-steady response, while larger values indicate a stronger relative departure. We inspect $|P^+|$ and $|P^+_{\rm M}|$ alongside $\epsilon_P$ to assess the coherence strength associated with these departures.

The output-field Rabi frequency provides a complementary measure of interaction strength,
\begin{equation}
\Omega(\tau)=\frac{2d}{\hbar}\left|E^+(L,\tau)\right|.
\label{eq:rabi_frequency}
\end{equation}
The product $\Omega(\tau)T_2(\tau)$ is the ratio of the dephasing time $T_2$ to the characteristic field--molecule coupling time $\Omega^{-1}$ at the sample output. Values approaching or exceeding unity indicate that the field can drive coherent evolution on a timescale comparable to or shorter than $T_2$. A strong field can also support a quasi-steady response, so $\Omega T_2$ alone does not determine the radiation regime.

To summarise these time-dependent diagnostics for each modelled flare, we calculate the polarisation-amplitude-weighted mean $\langle\epsilon_P\rangle_{\rm flare}$ and the maximum $(\Omega T_2)_{\max}$ over its half-maximum interval $[\tau_a,\tau_b]$. Table~\ref{tab:regime} reports the median and range of these quantities across the flares of each fitted solution. The local baseline, window definition, floor, weighting, and numerical checks are specified in \ref{app:diagnostics}.

We assign three operational labels using the solution median of the per-flare $\langle\epsilon_P\rangle_{\rm flare}$: values below 0.10 are labelled \emph{maser-tending}, values from 0.10 to below 0.25 \emph{superradiance-tending}, and values of 0.25 or above \emph{superradiance}. These empirical divisions lie in gaps between the groups represented in the fitted sample and do not define universal physical thresholds. The labels summarise each fitted interval, within which individual flares can have different departures. The complementary diagnostic $(\Omega T_2)_{\max}$ characterises field interaction relative to dephasing and does not independently determine the classification or saturation state.

The quasi-steady maser limit corresponds to $P^+\simeq P^+_{\rm M}$ and hence $\epsilon_P\simeq0$ while appreciable polarisation is present; it does not require $\Omega T_2\ll1$. No fitted solution is assigned a separate \emph{maser} label.

\section{Results}
\label{sec:results}

\subsection{Light-curve fits}
\label{sec:response}

\begin{figure}
\centering
\includegraphics[width=\linewidth]{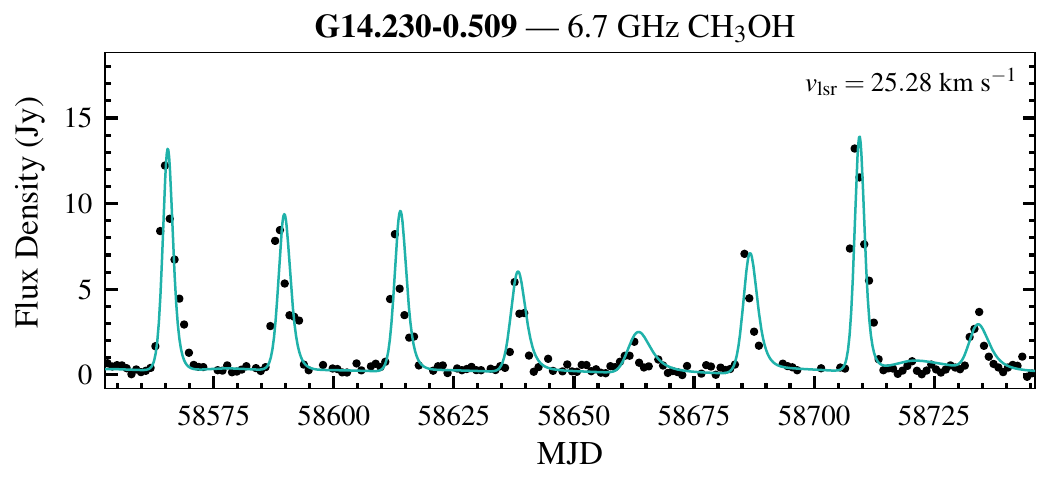}
\caption{Light curve of the $v_{\rm lsr}=25.28~\mathrm{km\,s^{-1}}$ component in G14.230$-$0.509 (black points) and its representative Maxwell--Bloch solution (cyan). The symmetric pump pulses have a period of 23.89~d, a width $\Delta t_{p,1/2}=2.19$~d, and amplitudes that vary between cycles.}
\label{fig:g14}
\end{figure}

\begin{figure}
\centering
\includegraphics[width=\linewidth]{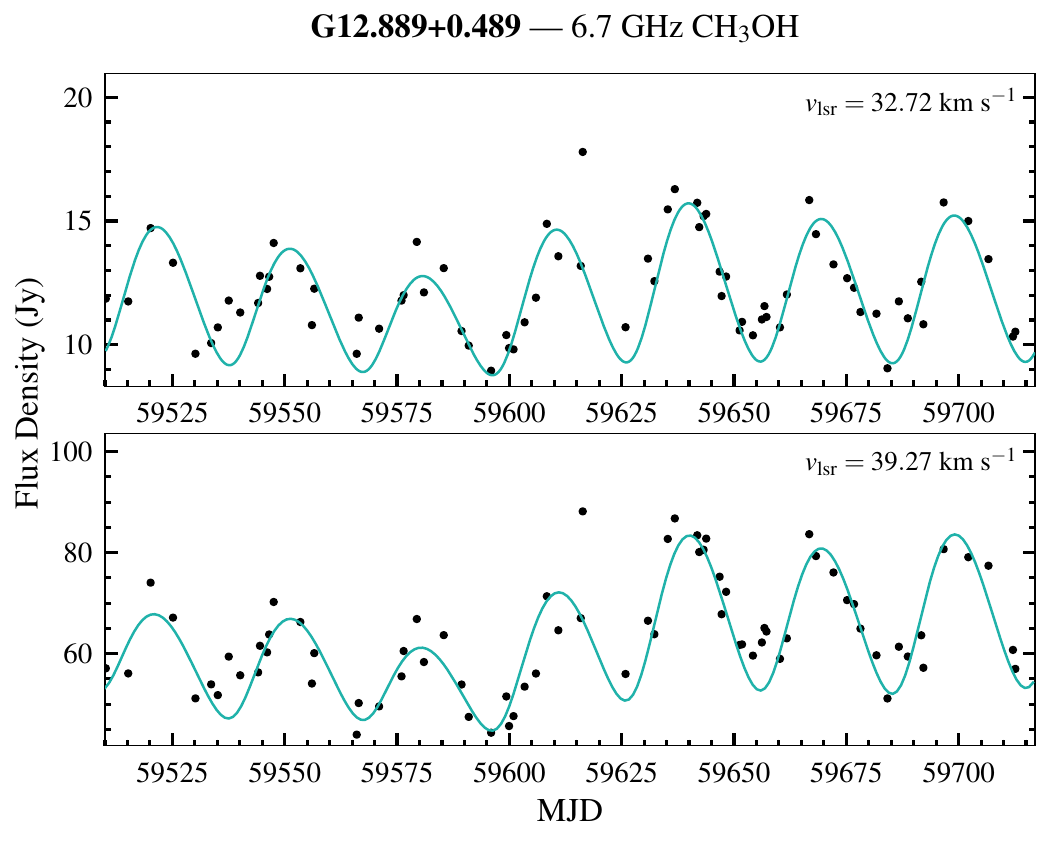}
\caption{Selected light curves of the $v_{\rm lsr}=32.72$ and $39.27~\mathrm{km\,s^{-1}}$ components in G12.889$+$0.489 (black points) and their representative Maxwell--Bloch solutions (cyan). The component periods are fixed to their measured values near 29.6~d, and both solutions use symmetric pump pulses with $\Delta t_{p,1/2}=7.05$~d.}
\label{fig:g12}
\end{figure}

\begin{figure}
\centering
\includegraphics[width=\linewidth]{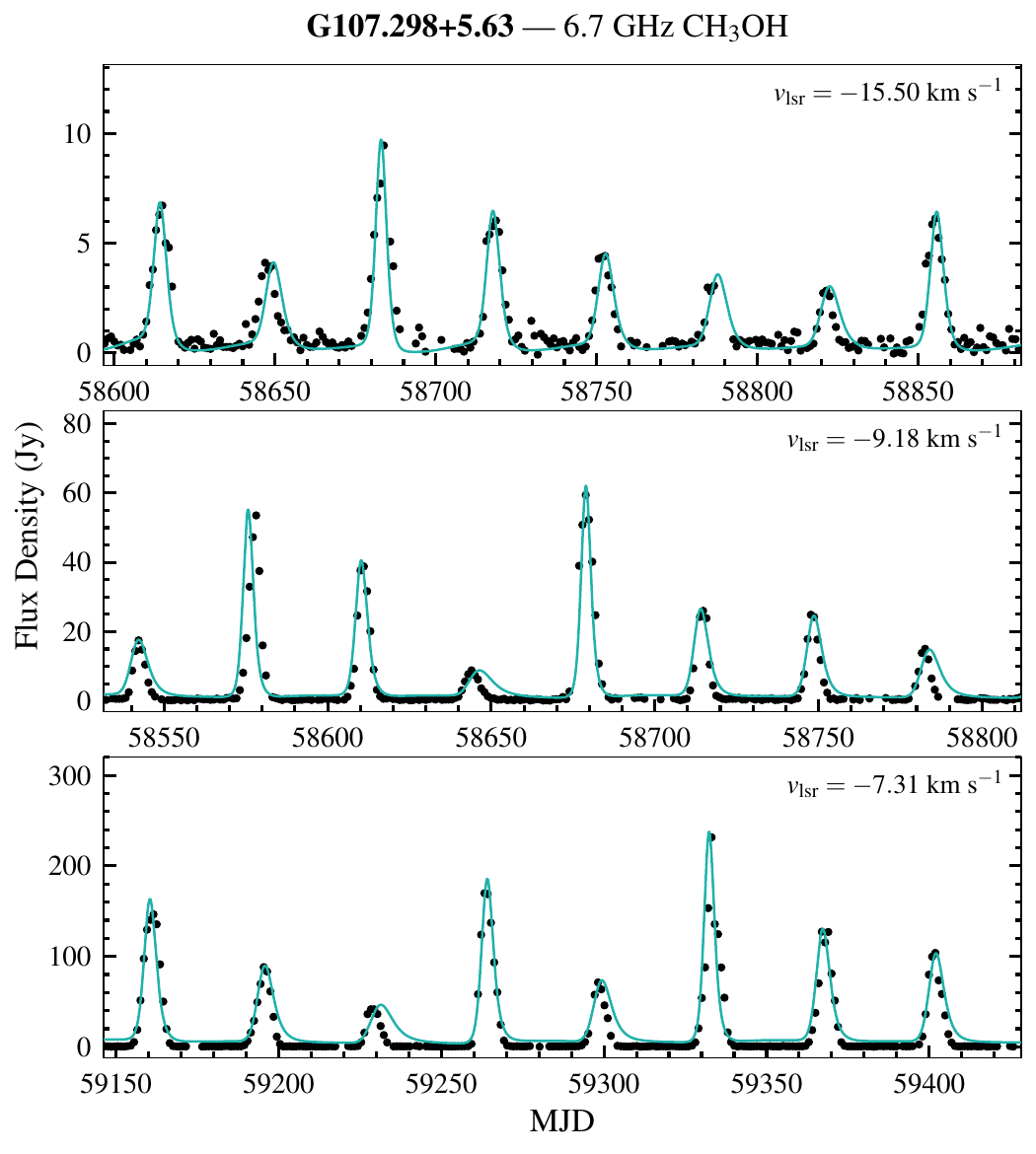}
\caption{Selected 6.7~GHz methanol light curves of three velocity components in G107.298$+$5.63 (black points) and their representative Maxwell--Bloch solutions (cyan). The component periods are fixed to their measured values near 34.4~d, and all three solutions use symmetric pump pulses with $\Delta t_{p,1/2}=3.47$~d.}
\label{fig:g107}
\end{figure}

\begin{figure}
\centering
\includegraphics[width=\linewidth]{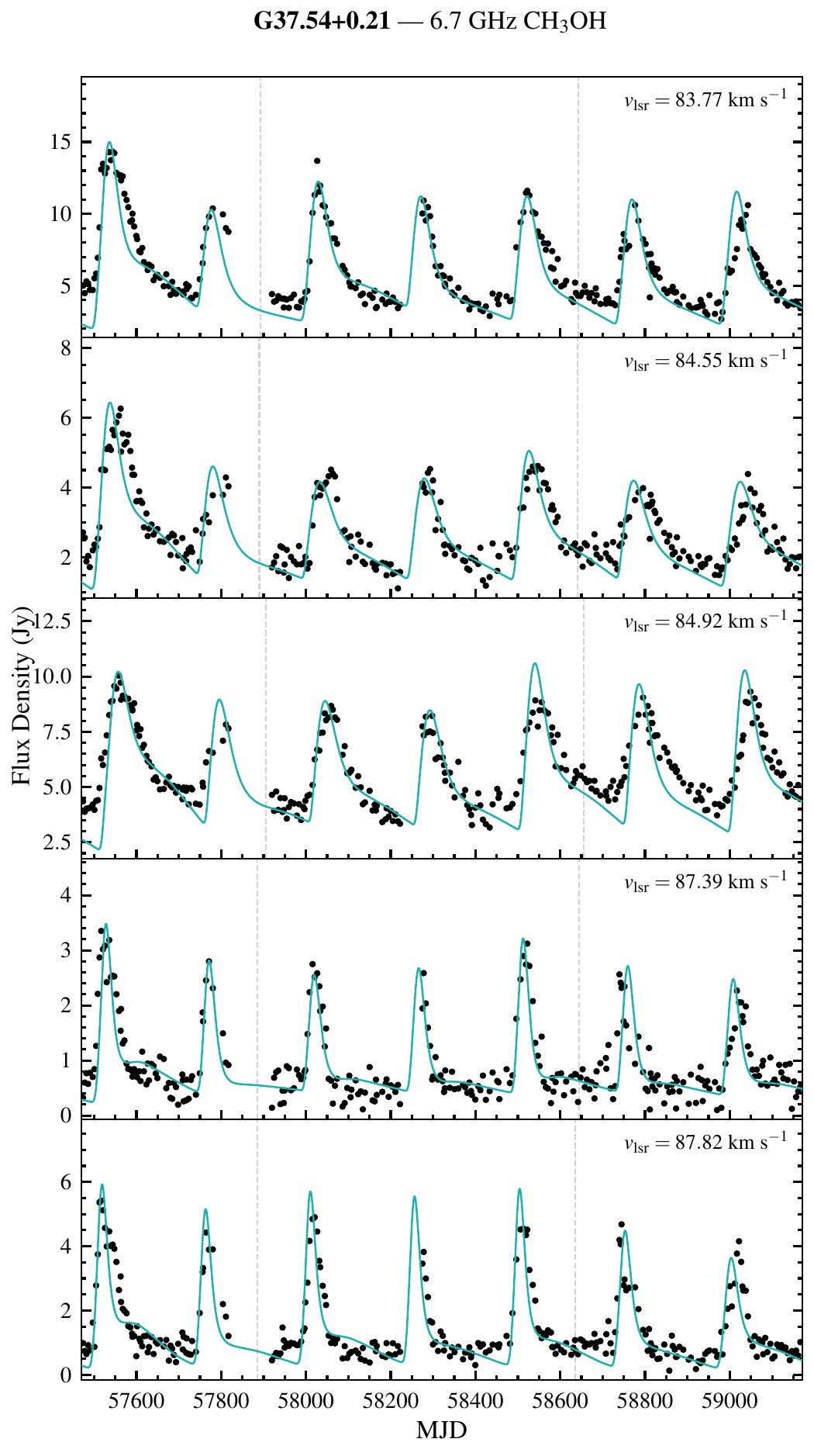}
\caption{Selected 6.7~GHz methanol light curves of five velocity components in G37.54$+$0.21 (black points) and their representative Maxwell--Bloch solutions (cyan). The component periods are fixed to their measured values near 247~d, and all five solutions use symmetric pump pulses with $\Delta t_{p,1/2}=7.05$~d. Each component is represented by one continuous solution with smoothly varying $T_2$; vertical dashed lines mark the centres of the two transitions in Equation~\eqref{eq:t2_steps}.}
\label{fig:g37}
\end{figure}

With symmetric pump pulses, the representative Maxwell--Bloch solutions reproduce the main fast-rise/slow-decay and approximately symmetric flare profiles in the selected, undetrended light curves (Figures~\ref{fig:g14}--\ref{fig:g37}). Cycle-to-cycle changes in flare strength are accommodated by the fitted pulse amplitudes $\Lambda_{1,m}$. Table~\ref{tab:params} lists the fitted intervals and parameters, while Table~\ref{tab:timescale_summary} compares the observed folded-profile widths with the imposed pump-pulse widths and model timescales.

The observed folded-profile widths of the modelled components exceed the imposed pump widths. The G12.889$+$0.489 and G37.54$+$0.21 fits share the same symmetric pump width, $\Delta t_{p,1/2}=7.05$~d, and have comparable initial superradiance timescales, $T_{\rm R,0}=0.59$--$0.69$~d. Nevertheless, the modelled G12.889$+$0.489 components have approximately symmetric folded profiles ($R_{\rm r/f}\simeq1$) with $\Delta t_{1/2}=11.6$--$12.3$~d, whereas the modelled G37.54$+$0.21 components have fast-rise/slow-decay profiles ($R_{\rm r/f}=0.30$--$0.47$) with $\Delta t_{1/2}=45.4$--$81.9$~d.

The narrower, high-contrast G14.230$-$0.509 and G107.298$+$5.63 responses are fitted with initial inverted column densities $n_0L$ roughly 2--4 times those of the broader G12.889$+$0.489 and G37.54$+$0.21 responses, giving shorter $T_{\rm R,0}$. Their dephasing times are also shorter. G14.230$-$0.509 and G107.298$+$5.63 nevertheless have the shortest and longest fitted population-relaxation times, respectively, with the broader responses lying between them. All solutions satisfy Equation~\eqref{eq:timescale_condition}, with $T_2<T_1$.

For G37.54$+$0.21, each continuous solution covers approximately 1700~d. The smooth evolution prescribed by Equation~\eqref{eq:t2_steps} increases $T_2$ by 0.7--1.5~d (approximately 6--11 per cent) near MJD~57890 and decreases it by 0.2--0.7~d near MJD~58640. The remaining medium parameters, pump width, and flux-density scale remain fixed within each fit.

\begin{table}[t]
\centering
\caption{Observed folded-profile widths $\Delta t_{1/2}$, imposed pump-pulse widths $\Delta t_{p,1/2}$, and model timescales for the components included in the Maxwell--Bloch fits. Ranges span the modelled components, while $\Delta t_{p,1/2}$ is common to the modelled components within each source. The $T_2$ range for G37.54$+$0.21 includes all three plateau values of each component. Width uncertainties and individual fit parameters are given in Tables~\ref{tab:metrics} and~\ref{tab:params}, respectively.}
\label{tab:timescale_summary}

\scriptsize
\setlength{\tabcolsep}{4pt}
\renewcommand{\arraystretch}{1.2}

\begin{tabular}{@{}lccccc@{}}
\toprule
Source
& $\Delta t_{1/2}$
& $\Delta t_{p,1/2}$
& $T_{\rm R,0}$
& $T_2$
& $T_1$ \\
& (d) & (d) & (d) & (d) & (d) \\
\midrule
G12.889$+$0.489
& 11.63--12.32
& 7.05
& 0.62--0.67
& 10.4--11.3
& 155--159 \\

G14.230$-$0.509
& 3.32
& 2.19
& 0.25
& 2.31
& 69 \\

G37.54$+$0.21
& 45.36--81.89
& 7.05
& 0.59--0.69
& 11.1--14.7
& 217--248 \\

G107.298$+$5.63
& 4.61--5.38
& 3.47
& 0.16--0.21
& 3.10--5.51
& 300 \\
\bottomrule
\end{tabular}
\end{table}

\subsection{Coherence-diagnostic results}
\label{sec:regime_results}

\begin{figure}
\centering
\includegraphics[width=\linewidth]{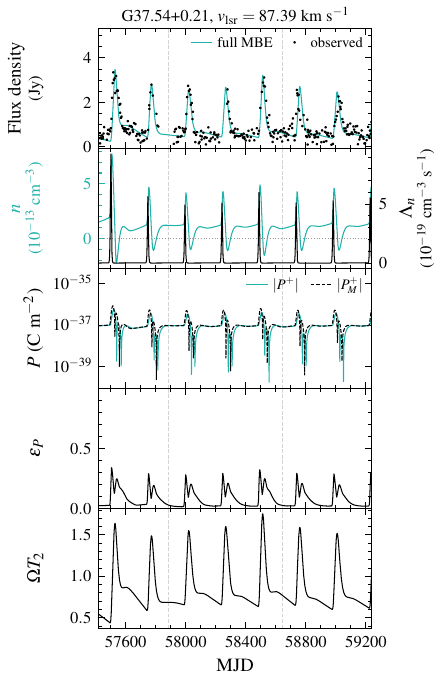}
\caption{Maxwell--Bloch solution and coherence diagnostics for the $v_{\rm lsr}=87.39~\mathrm{km\,s^{-1}}$ component of G37.54$+$0.21. From top to bottom: observed (black points) and modelled (cyan) flux density; output inversion density $n(L,\tau)$ (cyan, left axis) and total inversion-pump rate $\Lambda_n(\tau)$ (black, right axis); output polarisation magnitudes $|P^+|$ (cyan) and $|P^+_{\rm M}|$ (black dashed); column-integrated departure $\epsilon_P$; and output-field $\Omega(\tau)T_2(\tau)$. Here $P^+_{\rm M}$ is the instantaneous quasi-steady reference defined in Equation~\eqref{eq:PM}. Vertical dashed lines mark the centres of the two smooth $T_2$ transitions.}

\label{fig:g37_diagnostics}
\end{figure}

We examine the polarisation dynamics underlying the fitted light curves using the diagnostics defined in Section~\ref{sec:diagnostics}. Table~\ref{tab:regime} summarises the per-flare quantities for the eleven solutions, with their time dependence shown in Figure~\ref{fig:g37_diagnostics} and Figures~\ref{fig:diag_g12_pair}--\ref{fig:diag_g37_four}.

The five fitted G37.54$+$0.21 components share fast-rise/slow-decay profiles and a common imposed pump shape and width, yet occupy two diagnostic groups. The 83.77, 84.55, and $84.92~\mathrm{km\,s^{-1}}$ components are maser-tending, with median $\langle\epsilon_P\rangle_{\rm flare}$ values of 0.051--0.082. The 87.39 and $87.82~\mathrm{km\,s^{-1}}$ components have larger median departures, 0.160--0.168, and are superradiance-tending. A shared fast-rise/slow-decay morphology therefore does not uniquely identify the coherence dynamics of the fitted response.

The two G12.889$+$0.489 solutions are also maser-tending, with median $\langle\epsilon_P\rangle_{\rm flare}$ values of 0.066--0.077, despite median $(\Omega T_2)_{\max}$ values of 1.24--1.53. The two diagnostics describe different aspects of the response: $\langle\epsilon_P\rangle_{\rm flare}$ measures the realised departure from quasi-steady polarisation, whereas $(\Omega T_2)_{\max}$ measures the strength of field action relative to dephasing. The G12.889$+$0.489 solutions therefore remain close to the quasi-steady limit even though $\Omega T_2$ exceeds unity, illustrating why the operational labels are assigned from $\langle\epsilon_P\rangle_{\rm flare}$ rather than from the field ratio alone.

The largest median departures occur in G14.230$-$0.509 and the $-15.50~\mathrm{km\,s^{-1}}$ component of G107.298$+$5.63. Their median $\langle\epsilon_P\rangle_{\rm flare}$ values, 0.285 and 0.359, place both in the superradiance group; the corresponding median $(\Omega T_2)_{\max}$ values are 2.48 and 3.49. The other two fitted G107.298$+$5.63 components have smaller median departures, near 0.21, and are superradiance-tending.

Within the G37.54$+$0.21 solutions, the per-flare diagnostics vary across the long fitted intervals, including during the smooth evolution of $T_2$, but the operational labels in Table~\ref{tab:regime} summarise the median behaviour of each continuous solution rather than that of individual cycles.

\begin{table*}[t]
\centering
\caption{Coherence diagnostics and operational response labels for the eleven Maxwell--Bloch solutions. Values are medians of the per-flare diagnostics, with minimum--maximum ranges in parentheses. Labels are assigned from the solution median of $\langle\epsilon_P\rangle_{\rm flare}$ using the sample-specific criteria in Section~\ref{sec:diagnostics}. The statistics use seven flares per solution for G12.889$+$0.489 and G37.54$+$0.21, and eight for G14.230$-$0.509 and G107.298$+$5.63. Each G37.54$+$0.21 row covers one continuous solution with time-dependent $T_2$.}
\label{tab:regime}

\footnotesize
\setlength{\tabcolsep}{6pt}
\renewcommand{\arraystretch}{1.15}

\begin{tabular}{@{}lcccl@{}}
\toprule
Source & $v_{\rm lsr}$ & $\langle\epsilon_P\rangle_{\rm flare}$ & $(\Omega T_2)_{\max}$ & Response label \\
& ($\mathrm{km\,s^{-1}}$) & & & \\
\midrule

\multirow{2}{*}{G12.889$+$0.489}
& 32.72 & 0.077 (0.062--0.084) & 1.53 (1.42--1.58) & Maser-tending \\
& 39.27 & 0.066 (0.049--0.074) & 1.24 (1.15--1.34) & Maser-tending \\

\midrule

G14.230$-$0.509
& 25.28 & 0.285 (0.142--0.361) & 2.48 (1.37--3.23) & Superradiance \\

\midrule

\multirow{5}{*}{G37.54$+$0.21}
& 83.77 & 0.082 (0.062--0.113) & 1.00 (0.91--1.09) & Maser-tending \\
& 84.55 & 0.051 (0.049--0.081) & 0.79 (0.78--0.93) & Maser-tending \\
& 84.92 & 0.056 (0.047--0.067) & 0.84 (0.79--0.91) & Maser-tending \\
& 87.39 & 0.160 (0.147--0.196) & 1.59 (1.49--1.75) & Superradiance-tending \\
& 87.82 & 0.168 (0.126--0.177) & 1.46 (1.17--1.53) & Superradiance-tending \\

\midrule

\multirow{3}{*}{G107.298$+$5.63}
& $-15.50$ & 0.359 (0.267--0.455) & 3.49 (2.60--4.66) & Superradiance \\
& $-9.18$ & 0.207 (0.107--0.337) & 1.96 (1.15--3.05) & Superradiance-tending \\
& $-7.31$ & 0.211 (0.125--0.313) & 2.00 (1.27--2.86) & Superradiance-tending \\

\bottomrule
\end{tabular}
\end{table*}
\section{Discussion}
\label{sec:discussion}

The representative fits show that the temporal profile of the emitted light curve can differ from that of the imposed excitation. Symmetric pump pulses reproduce fast-rise/slow-decay and approximately symmetric flare profiles, while components sharing an imposed pulse width have different durations and asymmetries. The observed folded-profile widths also exceed the adopted pump widths. These results support interpreting flare morphology through the coupled evolution of the population inversion, polarisation, and radiation field, rather than as a direct record of the excitation pump pulses.

The characteristic superradiance timescale remains useful for interpreting cooperative emission, but its role depends on the excitation history. For a freshly inverted sample, it provides a scale for the ensuing burst and its build-up \citep{Rajabi_2020_maser}. In periodically driven systems, however, $T_{\rm R,0}$ refers to the initial inverted column density, while subsequent pulses act on the state produced by earlier evolution \citep{Rajabi2023}. The initial ordering $T_{\rm R,0}<T_2$ therefore does not establish the regime of every recurrent flare or uniquely predict its duration or peak time relative to the pump. The coherence diagnostics extend these timescale arguments by assessing the polarisation dynamics within the fitted flares.

The relevant distinction is whether the polarisation remains close to its instantaneous quasi-steady value. Maser-like evolution can persist while the inversion remains time dependent \citep{Rajabi_2020_maser}, so differences between the pump and emission profiles do not uniquely identify superradiance. The diagnostic $\epsilon_P$ directly measures departure from this polarisation approximation. By comparison, $\Omega T_2$ measures field coupling relative to dephasing: it can exceed unity while the driving and dephasing terms remain nearly balanced, leaving a small net polarisation derivative. This explains how the G12.889$+$0.489 solutions can be maser-tending despite their field ratios exceeding unity.

Although all fitted solutions satisfy the initial timescale ordering, their coherence diagnostics span maser-tending to superradiance behaviour. The narrow, high-contrast fitted responses in G14.230$-$0.509 and G107.298$+$5.63 occupy the superradiance and superradiance-tending groups, whereas the two G12.889$+$0.489 solutions are maser-tending. In G37.54$+$0.21, the fitted components share fast-rise/slow-decay profiles and a common pump shape and width, yet include both maser-tending and superradiance-tending responses. Thus, even within one source, a shared broad morphology does not uniquely specify the polarisation dynamics. The operational labels summarise the median per-flare coherence departure within each fitted interval, while individual flares can differ. They do not define universal boundaries between radiation regimes.

The fitted medium parameters show a related comparative pattern. Narrower responses have larger initial inverted column densities and shorter $T_{\rm R,0}$ and $T_2$, whereas their population-relaxation times occupy both ends of the fitted range (Table~\ref{tab:params}). Flare duration therefore cannot be attributed to a single relaxation or cooperative timescale. It depends on their combined effects, the excitation strength and repetition period, and the evolving state of the medium.

For G37.54$+$0.21, the constant medium parameters explored did not reproduce the full selected intervals, and the adopted solutions use modest, smooth changes in $T_2$. These changes occur across the multi-cycle interval and are not repeated with each pump pulse. With a period near 247~d, the fitted intervals span approximately 1700~d, compared with approximately 200--300~d for the other sources. This longer span may expose gradual evolution that the shorter fits do not sample. Within the adopted model, the fitted variation in $T_2$ describes the long-term profile evolution through changes in effective dephasing; the physical origin of these changes remains undetermined.

The inversion pump is also phenomenological. We adopt a symmetric pulse shape that is not independently constrained by simultaneous infrared monitoring. The reported correlation between formaldehyde and methanol flaring in G37.54$+$0.21 is consistent with a common radiative influence, but does not determine its origin or temporal profile \citep{Araya2010}. The present fits therefore constrain possible responses to excitation without identifying the astrophysical mechanism responsible for the periodicity.

Within the adopted geometry and excitation prescription, we obtain representative fits by manually adjusting the medium parameters, guided by visual agreement and the unweighted sum of squared residuals. Other parameter combinations may reproduce the light curves comparably well. In particular, the observations constrain the initial inverted column $n_0L$ more directly than $n_0$ and $L$ separately. The fitted inverted column densities and relaxation and dephasing times provide useful comparisons of the effective conditions associated with different flare profiles.

Strongly slow-rise/fast-decay profiles, including the $79.62~\mathrm{km\,s^{-1}}$ component of G37.54$+$0.21, remain outside the fitted subset because acceptable solutions with one effective sample were not obtained within the parameter searches performed.

Applying the same diagnostics to earlier periodic Maxwell--Bloch solutions would extend the comparison of response regimes across sources. Simultaneous infrared and maser monitoring would constrain the excitation history, while time-resolved VLBI observations could constrain the spatial distribution and variability of the emitting regions.

\section{Conclusions}
\label{sec:conclusions}

We analysed periodic 6.7~GHz methanol flares in four sources with approximately stable quiescent emission. Their folded profiles span a broad range of fractional widths, contrasts, and rise-to-fall ratios, showing that components with similar repetition periods can have markedly different morphologies.

Eleven representative Maxwell--Bloch solutions reproduce selected fast-rise/slow-decay and approximately symmetric light curves using a common symmetric pump shape and width within each source, with component-dependent medium parameters and cycle-dependent pulse amplitudes. The observed folded-profile widths exceed the imposed pump widths, and components sharing a pump pulse width have different durations and asymmetries. The G37.54$+$0.21 fits also include smooth changes in $T_2$ across the multi-cycle intervals.

The coherence diagnostics extend the characteristic-timescale analysis by assessing the polarisation dynamics realised during the fitted periodic flares. In these continuously driven solutions, $T_{\rm R,0}$ characterises the initial inverted column, while successive pulses act on the evolving state of the medium. The fitted responses span three operational groups: maser-tending, superradiance-tending, and superradiance. In G37.54$+$0.21, components with similar fast-rise/slow-decay profiles occupy different groups, illustrating the additional information provided by the coherence dynamics.

Acceptable fits to strongly slow-rise/fast-decay profiles were not obtained with one effective sample within the parameter searches performed. For the modelled profiles, the fits constrain effective medium conditions under the adopted excitation prescription. The excitation pump and evolving medium jointly shape the emission, while the coherence diagnostics distinguish the dynamics underlying the fitted light curves.

\section*{Acknowledgments}

F.R. acknowledges support from the NSERC Discovery Grant RGPIN-2024-06346. This work was also partially supported by the Inter-university Collaborative Project ``Japanese VLBI Network (JVN)'' of NAOJ and the Japan Society for the Promotion of Science (JSPS) KAKENHI Grants JP21H01120 and JP21H00032. A.B. acknowledges support from the National Science Centre, Poland, under Grant No. 2021/43/B/ST9/02008.

\section*{Declaration of generative AI and AI-assisted technologies in the manuscript preparation process}

During the preparation of this manuscript, the authors used ChatGPT and Claude Opus~5.5 to assist with language refinement, clarity, and organisation. The authors reviewed and edited the output and take full responsibility for the content of the publication.

\appendix

\makeatletter
\@addtoreset{figure}{section}
\@addtoreset{table}{section}
\renewcommand{\thefigure}{\Alph{section}.\arabic{figure}}
\renewcommand{\thetable}{\Alph{section}.\arabic{table}}
\makeatother

\section{Period-search procedure and \texorpdfstring{G37.54$+$0.21}{G37.54+0.21} periodograms}
\label{app:periods}

Periods were measured with Lomb--Scargle (LS) periodograms \citep{Lomb1976,Scargle1982}. For G12.889$+$0.489 and G37.54$+$0.21, long-term baseline variations were removed using the asymmetrically reweighted penalised least-squares method of \citet{Baek2015}, with smoothing parameters $\lambda=10^9$ and $10^{10}$, respectively. G14.230$-$0.509 and G107.298$+$5.63 required no detrending.

The baseline was estimated by iteratively minimising
\[
\sum_i w_i(y_i-b_i)^2+\lambda\sum_i[b''(t_i)]^2,
\]
where $y_i$ is the flux density measured at time $t_i$, $b_i=b(t_i)$ is the fitted baseline, and $b''(t_i)$ is its second derivative with respect to time. The weights $w_i$ are updated from the residuals to downweight measurements well above the baseline. The roughness penalty uses the actual observing times to account for uneven sampling. We adopted a weight-convergence tolerance of $10^{-3}$ and a maximum of 50 iterations. The detrended values are $y_i^{\rm det}=y_i-b_i+\bar b$, where $\bar b$ is the mean fitted baseline. Detrending was used only for the period search; the original flux densities were retained for the folded-profile measurements and Maxwell--Bloch modelling.

The LS periodograms were calculated with \texttt{astropy.timeseries.LombScargle}, using standard normalisation and a frequency grid constructed with \texttt{samples\_per\_peak=20}, covering trial periods of 10--1000~d. Peak significance was evaluated separately for each velocity component using 1000 Monte Carlo realisations sampled at the observed epochs. For each realisation, the observed epochs were retained and the flux densities replaced by uncorrelated Gaussian random numbers of zero mean and unit variance. Each realisation was processed through the same period-search pipeline as the corresponding observed light curve. The 99th percentile of the maximum simulated LS powers defined the empirical detection threshold under this Gaussian-noise null model.

The half-width at half maximum (HWHM) of each accepted LS peak is used as a measure of period resolution, not as a formal statistical uncertainty. For sources with multiple periodic components, the representative period in Table~\ref{tab:periods} is the weighted mean of the component periods, with weights $w_k=(\Delta p_{{\rm HWHM},k})^{-2}$. We use this mean only as a descriptive summary and assign it no formal uncertainty, since the component periods are not assumed to be statistically independent. The component periods adopted for folding are listed in Table~\ref{tab:metrics}.

\begin{figure}
\centering
\includegraphics[width=\linewidth]{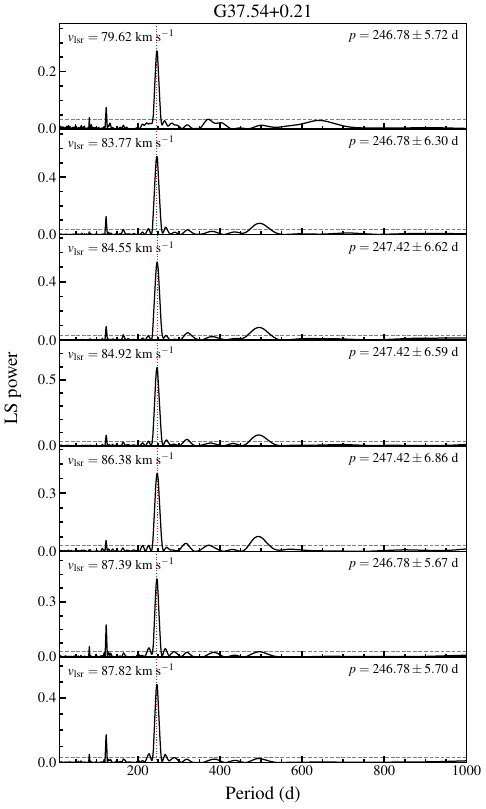}
\caption{Lomb--Scargle periodograms of the accepted periodic 6.7~GHz methanol components of G37.54$+$0.21. Labels give $v_{\rm lsr}$ and the peak period $\pm$ its LS-peak HWHM, used here as a resolution measure rather than a statistical uncertainty. Horizontal dashed lines mark the component-specific 99th-percentile Monte Carlo thresholds, and vertical dotted lines mark the accepted peaks near 247~d.}
\label{fig:g37_periodograms}
\end{figure}

Figure~\ref{fig:g37_periodograms} shows the LS peaks of the accepted G37.54$+$0.21 components near the representative period of 247~d.

\section{Folded-profile measurements}
\label{app:metrics}

The morphology quantities in Table~\ref{tab:metrics} are obtained by fitting an empirical split Gaussian to each representative folded profile. This fit describes the recurrent flare shape independently of the physical Maxwell--Bloch response model.

We fold the original flux densities over the MJD interval listed in Table~\ref{tab:metrics}, using the component period $p$ and
\[
N_{\rm bin}=\min[\operatorname{round}(p),35],
\]
with $p$ expressed in days. Within each phase bin, measurements are first averaged within individual cycles and then across cycles, giving each contributing cycle equal weight. The phase origin is placed at the midpoint of the quiescent interval of the binned profile so that the flare is not divided by the folding boundary.

The error bars on the binned profile describe the uncertainty in the mean flux density at each phase. For phase bin $i$, we use the standard error
\[
\sigma_i=\frac{s_i}{\sqrt{n_i}},
\]
where $n_i$ is the number of contributing cycles and $s_i$ is the sample standard deviation of the flux-density values contributed by those cycles. Thus, $s_i$ measures the cycle-to-cycle scatter, while $\sigma_i$ estimates the uncertainty in the phase-bin mean.

The fitted profile is
\[
S(t)=F_{\rm q}+A_{\rm fl}\exp\!\left[-\frac{(t-t_{\rm pk})^2}{2\sigma_k^2}\right],
\qquad
k=
\begin{cases}
{\rm r}, & t\leq t_{\rm pk},\\
{\rm f}, & t>t_{\rm pk},
\end{cases}
\]
where $t$ is the time within the folded cycle. The branches share the quiescent level $F_{\rm q}$, flare amplitude $A_{\rm fl}$, and peak time $t_{\rm pk}$, but have separate rising and falling widths $\sigma_{\rm r}$ and $\sigma_{\rm f}$. The full width at half maximum above $F_{\rm q}$ and the rise-to-fall ratio are
\[
\Delta t_{1/2}=(\sigma_{\rm r}+\sigma_{\rm f})\sqrt{2\ln2},
\qquad
R_{\rm r/f}=\frac{\sigma_{\rm r}}{\sigma_{\rm f}}.
\]

All populated phase bins receive equal weight in the fit; the displayed uncertainties $\sigma_i$ are not used as fitting weights. The estimated parameter covariance matrix is scaled by ${\rm RSS}/\nu$, where ${\rm RSS}$ is the residual sum of squares and $\nu$ is the number of residual degrees of freedom.

To propagate uncertainties while accounting for parameter correlations, we draw 300 sets of $(F_{\rm q},A_{\rm fl},t_{\rm pk},\sigma_{\rm r},\sigma_{\rm f})$ jointly from a multivariate normal distribution centred on the best-fitting values and using this covariance matrix. Draws with non-positive $A_{\rm fl}$, $\sigma_{\rm r}$, or $\sigma_{\rm f}$ are discarded. For each retained draw, we calculate the morphology quantities, including the fractional width $\Delta t_{1/2}/p$ and contrast $C_{\rm fl}=A_{\rm fl}/F_{\rm q}$, with $p$ held fixed. The corresponding 16th and 84th percentiles define the lower and upper uncertainty bounds. These uncertainties are conditional on the adopted period, phase binning, and MJD interval.

\begin{figure*}[p]
\centering
\includegraphics[width=\textwidth]{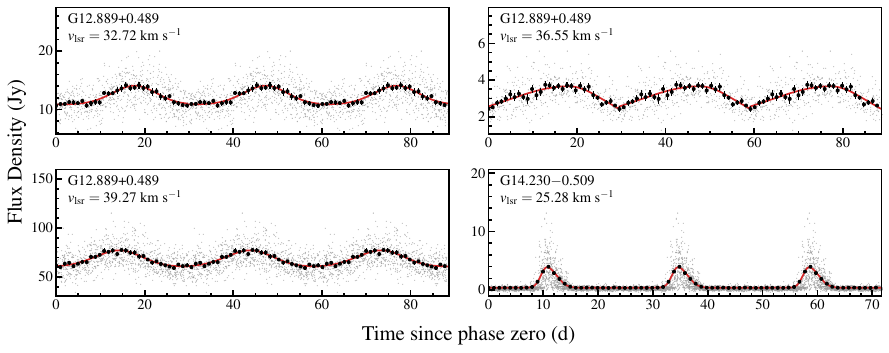}
\caption{Representative folded profiles of G12.889$+$0.489 and G14.230$-$0.509, repeated over three cycles for display. Grey points show individual measurements folded at the component period over the MJD interval in Table~\ref{tab:metrics}; those for G12.889$+$0.489 include both iMet and Toruń observations. Black points show phase-bin means obtained by averaging the cycle-level values, with error bars giving their standard errors. Red curves show the split-Gaussian fits. The phase origin lies at the midpoint of the quiescent interval.}
\label{fig_folded_g12_g14}
\end{figure*}

\begin{figure*}[p]
\centering
\includegraphics[width=\textwidth]{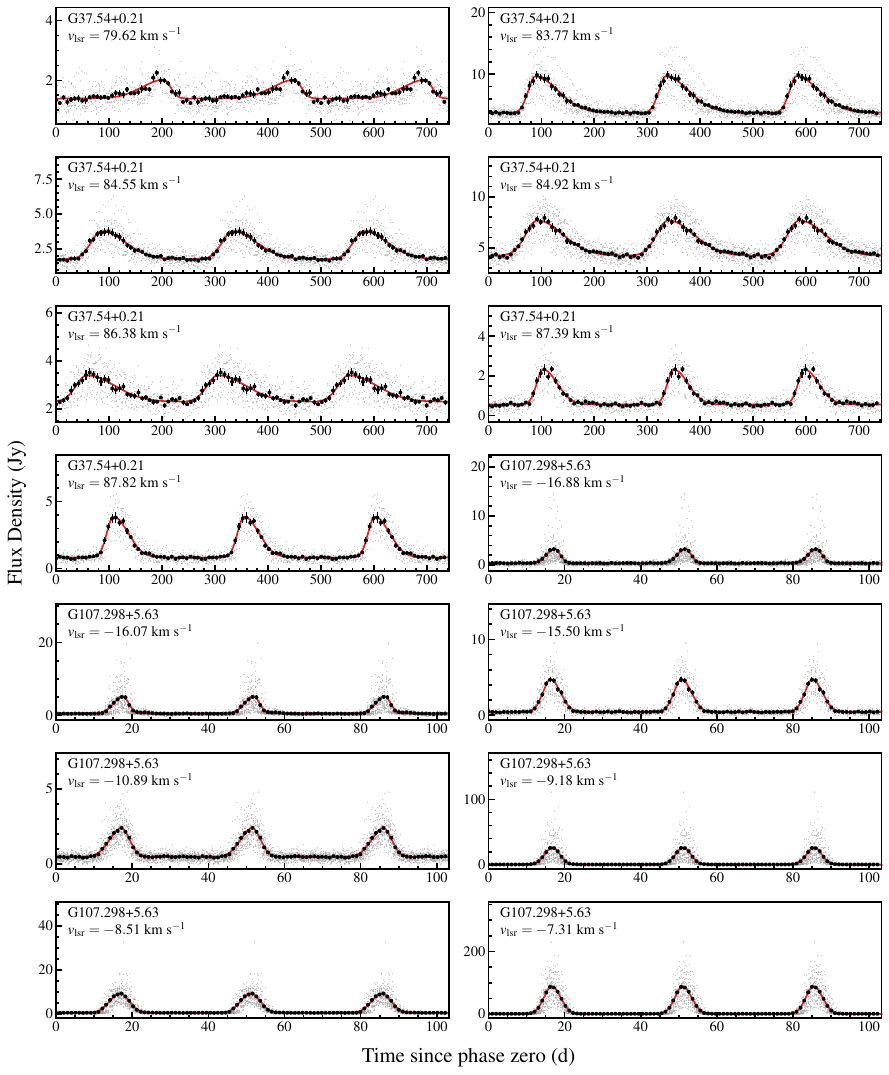}
\caption{Representative folded profiles of G37.54$+$0.21 and G107.298$+$5.63, repeated over three cycles for display. Symbols, error bars, and fitted curves follow the conventions of Figure~\ref{fig_folded_g12_g14}.}
\label{fig_folded_g37_g107}
\end{figure*}

Figures~\ref{fig_folded_g12_g14} and~\ref{fig_folded_g37_g107} show the folded profiles and their split-Gaussian fits.

\section{Maxwell--Bloch implementation}
\label{app:MBE}

\subsection{Equations and conventions}

For each effective two-level sample, we evolve the population-inversion density $n=n_{\rm u}-n_{\rm l}$, macroscopic polarisation amplitudes $P^\pm$, and electric-field amplitudes $E^\pm$. Under the rotating-wave and slowly varying-envelope approximations, the one-dimensional Maxwell--Bloch equations in retarded time $\tau=t-z/c$ are \citep{Gross_Haroche_1982,Benedict_1996,Rajabi_2020_maser}
\begin{subequations}
\label{eq:MBE}
\begin{align}
\frac{\partial n}{\partial\tau}
&=\frac{2i}{\hbar}\left(P^+E^+-P^-E^-\right)
-\frac{n}{T_1}
+\Lambda_n(\tau),
\label{eq:MBE_n}\\
\frac{\partial P^+}{\partial\tau}
&=\frac{id^2}{\hbar}E^-n
-\frac{P^+-P^+_0}{T_2(\tau)},
\label{eq:MBE_P}\\
\frac{\partial E^+}{\partial z}
&=\frac{i\omega_0}{2\epsilon_0c}P^-.
\label{eq:MBE_field}
\end{align}
\end{subequations}
Here $d$ and $\omega_0$ are the transition dipole magnitude and angular frequency, respectively; $\epsilon_0$ and $c$ are the vacuum permittivity and speed of light. The amplitudes satisfy $P^-=(P^+)^*$ and $E^-=(E^+)^*$. Our electric-field convention is
\[
E_{\rm phys}(z,t)=E^+(z,\tau)e^{-i\omega_0\tau}+E^-(z,\tau)e^{i\omega_0\tau}.
\]

The inversion-pump rate $\Lambda_n(\tau)$ includes the constant background and periodic pulses defined in Equation~\eqref{eq:pump1}. Within each fit, $T_1$ is constant; $T_2$ follows Equation~\eqref{eq:t2_steps} for G37.54$+$0.21 and is constant otherwise.

For the seed prescription, we assume a spatially uniform initial state. The unit-Fresnel geometry $A=\lambda L$ gives
\[
N=n_0AL=n_0\lambda L^2,
\qquad
\theta_0=\frac{2}{\sqrt{N}},
\]
where $N$ is the effective number of emitters associated with the initial inversion and $\theta_0$ is the small initial Bloch angle representing the quantum fluctuation that seeds the emission \citep{Gross_Haroche_1982}. The initial state is
\[
n(z,0)=n_0\cos\theta_0,
\qquad
P^+(z,0)=P^+_0=\frac{n_0d}{2}\sin\theta_0.
\]
Here $n_0$ denotes the full population-inversion density $n_{\rm u}-n_{\rm l}$, and the factor of $1/2$ in $P^+_0$ follows from the positive-frequency polarisation convention adopted above. Thus, $n_0$ sets the initial inversion scale, with $n(z,0)\simeq n_0$ for $\theta_0\ll1$. The seed polarisation $P^+_0$ is retained as the fixed reference in the polarisation-relaxation term. The background pump is
\[
\Lambda_{n,0}=\frac{n_0\cos\theta_0}{T_1}\simeq\frac{n_0}{T_1}.
\]
All fits impose a zero incident-field boundary condition, $E^+(0,\tau)=0$, with the prescribed initial polarisation seeding the emission.

Setting $\partial_\tau P^+=0$ in Equation~\eqref{eq:MBE_P} gives the instantaneous maser-limit polarisation,
\[
P^+_{\rm M}=P^+_0+\frac{id^2T_2(\tau)}{\hbar}E^-n.
\]
With the field convention above, the output intensity and Rabi frequency at $z=L$ are
\[
I_{\rm out}(\tau)=2c\epsilon_0\left|E^+(L,\tau)\right|^2,
\qquad
\Omega(\tau)=\frac{2d}{\hbar}\left|E^+(L,\tau)\right|.
\]

The numerical integration uses a fourth-order Runge--Kutta scheme with component-dependent time steps of approximately 0.2--2~d and a spatial step $\Delta\zeta \simeq 5\times10^{-4}$, where $\zeta=z/L$. Each fitted interval is integrated continuously, without resetting the state between pump cycles or across the smooth $T_2$ transitions.

\subsection{Diagnostic averaging and numerical checks}
\label{app:diagnostics}

Each modelled flare is analysed in a window of duration $p$ centred on its model peak. The minimum model flux density within this window defines the local baseline. The diagnostic interval $[\tau_a,\tau_b]$ is determined directly from the model light curve, using the half-maximum crossings on either side of the peak measured above this baseline.

The polarisation floor in Equation~\eqref{eq:epsP} is
\[
\begin{aligned}
P_{\rm ref}&=\max_{z,\tau}\left(|P^+|+|P^+_{\rm M}|\right),\\
P_{\rm floor}&=\max\left(\eta P_{\rm ref},\kappa P_{\rm num}\right).
\end{aligned}
\]
The maximum defining $P_{\rm ref}$ is taken over the full sample, $0\leq z\leq L$, and the full one-period analysis window. Here $P_{\rm num}$ is the median of $|P^+_{\rm adopted}-P^+_{\rm refined}|$, evaluated at matching positions and times over the quiescent portions of the analysis windows. The refined solution uses twice the resolution in both time and position. Quiescent points are those for which the model flux above the local baseline is less than 10 per cent of the peak flux above that baseline. We adopt $\eta=10^{-6}$ and $\kappa=5$. The floor has units of polarisation and enters the column-integrated diagnostic as $LP_{\rm floor}$.

We weight the flare average by the column-integrated polarisation amplitude,
\[
W_P(\tau)=\int_0^L\left(|P^+|+|P^+_{\rm M}|\right)\,dz,
\]
so that times with negligible polarisation contribute little. The per-flare diagnostics are
\[
\begin{aligned}
\left\langle\epsilon_P\right\rangle_{\rm flare}
&=
\frac{\displaystyle\int_{\tau_a}^{\tau_b}W_P(\tau)\epsilon_P(\tau)\,d\tau}
{\displaystyle\int_{\tau_a}^{\tau_b}W_P(\tau)\,d\tau},\\
(\Omega T_2)_{\max}
&=
\max_{\tau_a\leq\tau\leq\tau_b}
\left[\Omega(\tau)T_2(\tau)\right].
\end{aligned}
\]
Both diagnostics use the same half-maximum interval, with the instantaneous $T_2(\tau)$ included for G37.54$+$0.21. Table~\ref{tab:regime} reports their medians and minimum--maximum ranges across the modelled flares of each solution.

Varying $\eta$ over $10^{-8}$--$10^{-4}$ and $\kappa$ over 3--10 changes each solution's median $\langle\epsilon_P\rangle_{\rm flare}$ by at most 1.3 per cent; $(\Omega T_2)_{\max}$ is independent of the floor. Refining both numerical grids by a factor of two changes the solution medians of $\langle\epsilon_P\rangle_{\rm flare}$ and $(\Omega T_2)_{\max}$ by at most 2.3 and 1.5 per cent, respectively.

\subsection{Principal fit parameters}
\label{app:params}

Table~\ref{tab:params} lists the fitted intervals and principal parameters of the Maxwell--Bloch solutions shown in Figures~\ref{fig:g14}--\ref{fig:g37}.

\begin{table*}[p]
\centering
\caption{Principal parameters of the representative Maxwell--Bloch fits. Here $\Delta t_{p,1/2}$ is the pump-pulse full width at half maximum, and $\Lambda_{n,0}=n_0\cos\theta_0/T_1\simeq n_0/T_1$ is the background inversion-pump rate. The listed $\Lambda_{1,m}/\Lambda_{n,0}$ ranges give the minimum and maximum pulse amplitudes relative to this background across the fitted cycles. For G37.54$+$0.21, arrows indicate the successive $T_2$ plateau values in Equation~\eqref{eq:t2_steps}. The two smooth transitions have a smoothing timescale $\delta_k=100$~d and component-dependent centres within MJD~57885--57906 and 58635--58656, respectively.}
\label{tab:params}

\scriptsize
\setlength{\tabcolsep}{4.5pt}
\renewcommand{\arraystretch}{1.10}

\begin{tabular}{@{}llcccccccc@{}}
\toprule
Source & $v_{\rm lsr}$ & Interval & $T_1$ & $T_2$ & $\Delta t_{p,1/2}$ & $\Lambda_{n,0}$ & $\Lambda_{1,m}/\Lambda_{n,0}$ & $n_0L$ & $T_{\rm R,0}$ \\
& ($\mathrm{km\,s^{-1}}$) & (MJD) & (d) & (d) & (d) & ($10^{-19}~\mathrm{cm^{-3}\,s^{-1}}$) & & ($10^3~\mathrm{cm^{-2}}$) & (d)  \\
\midrule

\multirow{2}{*}{G12.889$+$0.489}
& 32.72 & 59510--59717 & 159 & 11.3 & 7.05 & 2.08 & 3.1--4.0 & 4.86 & 0.62  \\
& 39.27 & 59510--59717 & 155 & 10.4 & 7.05 & 1.99 & 3.2--4.8 & 4.52 & 0.67  \\

\midrule

G14.230$-$0.509
& 25.28 & 58552--58746 & 69 & 2.31 & 2.19 & 7.92 & 15--46 & 11.9 & 0.25 \\

\midrule

\multirow{5}{*}{G37.54$+$0.21}
& 83.77 & 57470--59187 & 222 & $13.0\rightarrow13.8\rightarrow13.1$ & 7.05 & 0.53 & 6.5--13 & 4.59 & 0.66 \\
& 84.55 & 57470--59171 & 222 & $12.3\rightarrow13.0\rightarrow12.8$ & 7.05 & 0.56 & 5.7--10 & 4.86 & 0.62 \\
& 84.92 & 57470--59187 & 248 & $12.3\rightarrow13.0\rightarrow12.8$ & 7.05 & 0.53 & 5.2--9.3 & 5.08 & 0.59 \\
& 87.39 & 57470--59187 & 217 & $13.2\rightarrow14.7\rightarrow14.5$ & 7.05 & 0.52 & 9.5--18 & 4.39 & 0.69 \\
& 87.82 & 57464--59171 & 224 & $11.1\rightarrow11.8\rightarrow11.3$ & 7.05 & 0.51 & 17--24 & 4.43 & 0.68 \\

\midrule

\multirow{3}{*}{G107.298$+$5.63}
& $-15.50$ & 58597--58882 & 300 & 5.51 & 3.47 & 2.88 & 9.8--21 & 18.6 & 0.16 \\
& $-9.18$ & 58532--58812 & 300 & 3.10 & 3.47 & 2.25 & 13--47 & 14.6 & 0.21 \\
& $-7.31$ & 59146--59428 & 300 & 3.10 & 3.47 & 2.42 & 16--39 & 15.7 & 0.19 \\

\bottomrule
\end{tabular}
\end{table*}
\FloatBarrier

\onecolumn

\section{Additional coherence-diagnostic figures}
\label{app:diagnostic_plots}

\begin{figure}[!htp]
\centering
\includegraphics[width=0.40\textwidth,height=0.46\textheight,keepaspectratio]{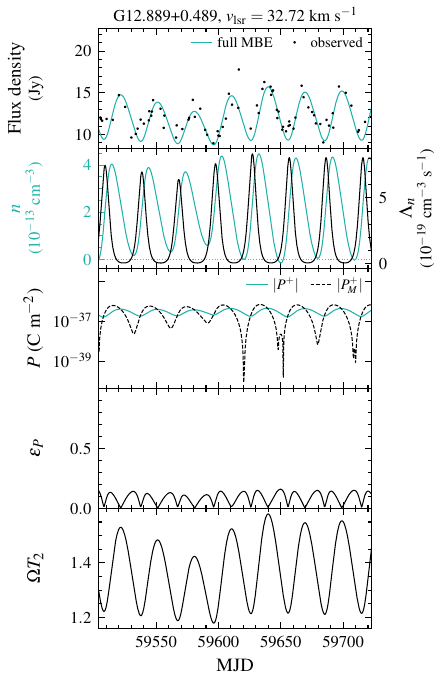}%
\hspace{0.05\textwidth}%
\includegraphics[width=0.40\textwidth,height=0.46\textheight,keepaspectratio]{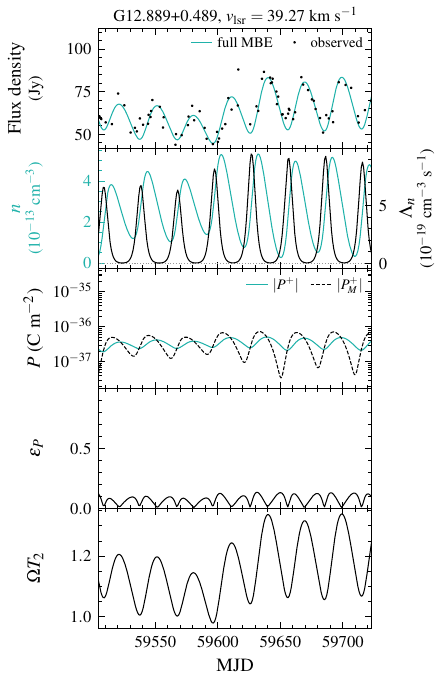}
\caption{Maxwell--Bloch solutions and coherence diagnostics for G12.889$+$0.489 at $v_{\rm lsr}=32.72$ (left) and $39.27~\mathrm{km\,s^{-1}}$ (right). In each plot, the panels show, from top to bottom: observed (black points) and modelled (cyan) flux density; output inversion density $n(L,\tau)$ (cyan, left axis) and total inversion-pump rate $\Lambda_n(\tau)$ (black, right axis); output polarisation magnitudes $|P^+|$ (cyan) and $|P^+_{\rm M}|$ (black dashed); column-integrated departure $\epsilon_P$; and output-field $\Omega(\tau)T_2(\tau)$. Here $P^+_{\rm M}$ is the instantaneous quasi-steady reference defined in Equation~\eqref{eq:PM}.}
\label{fig:diag_g12_pair}
\end{figure}

\begin{figure}[p]
\centering
\includegraphics[width=0.40\textwidth,height=0.46\textheight,keepaspectratio]{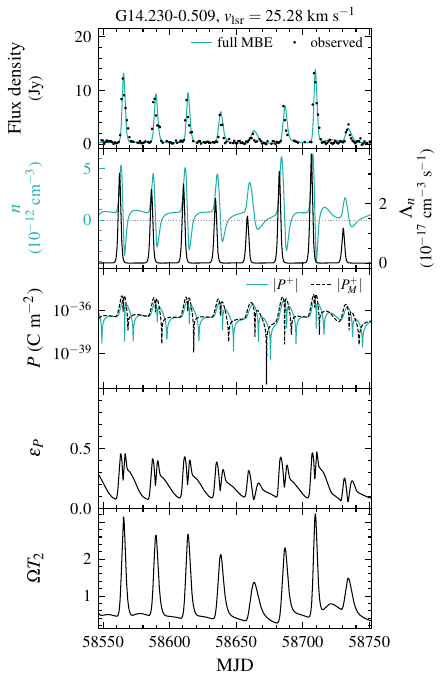}%
\hspace{0.05\textwidth}%
\includegraphics[width=0.40\textwidth,height=0.46\textheight,keepaspectratio]{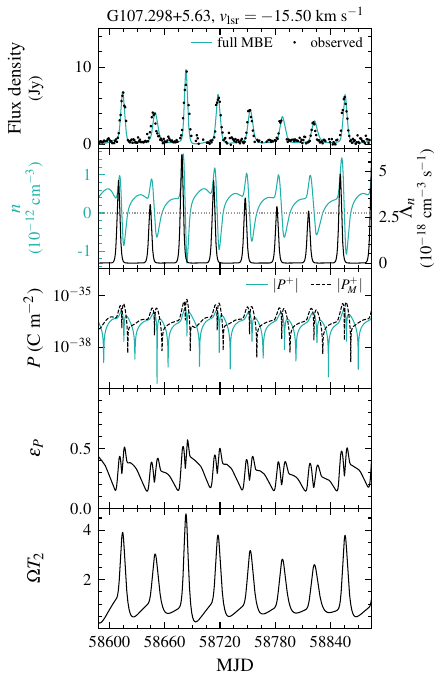}\\[2mm]
\includegraphics[width=0.40\textwidth,height=0.46\textheight,keepaspectratio]{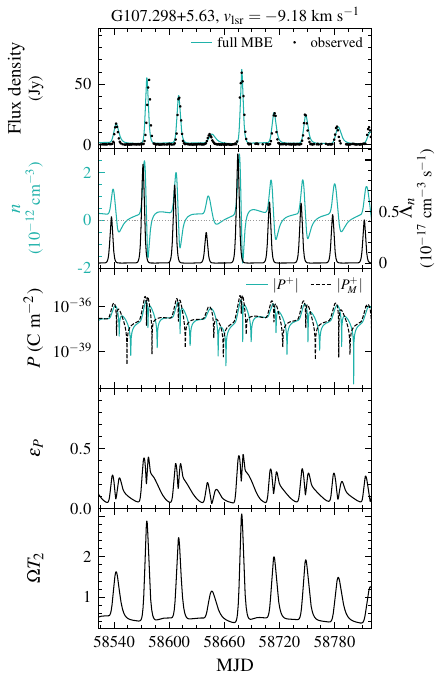}%
\hspace{0.05\textwidth}%
\includegraphics[width=0.40\textwidth,height=0.46\textheight,keepaspectratio]{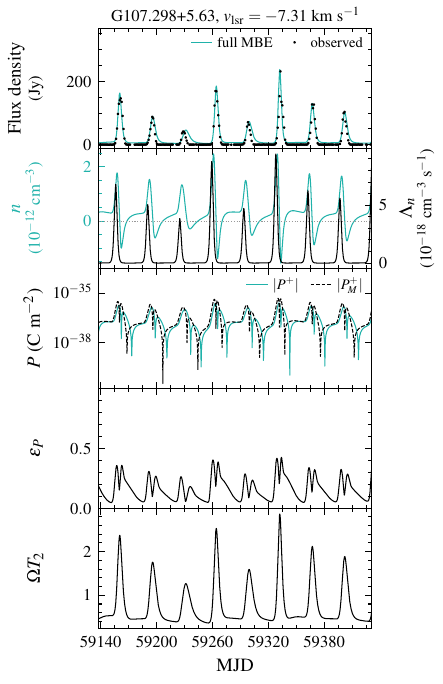}
\caption{Same as Figure~\ref{fig:diag_g12_pair}, for G14.230$-$0.509 at $v_{\rm lsr}=25.28~\mathrm{km\,s^{-1}}$ (top left) and G107.298$+$5.63 at $v_{\rm lsr}=-15.50$ (top right), $-9.18$ (bottom left), and $-7.31~\mathrm{km\,s^{-1}}$ (bottom right).}
\label{fig:diag_g14_g107}
\end{figure}

\begin{figure}[p]
\centering
\includegraphics[width=0.40\textwidth,height=0.46\textheight,keepaspectratio]{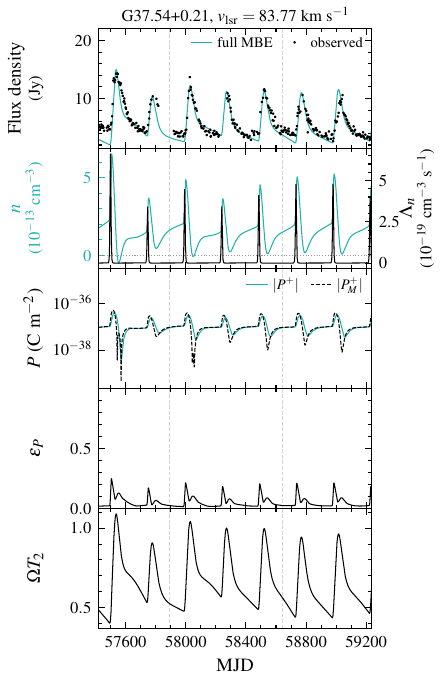}%
\hspace{0.05\textwidth}%
\includegraphics[width=0.40\textwidth,height=0.46\textheight,keepaspectratio]{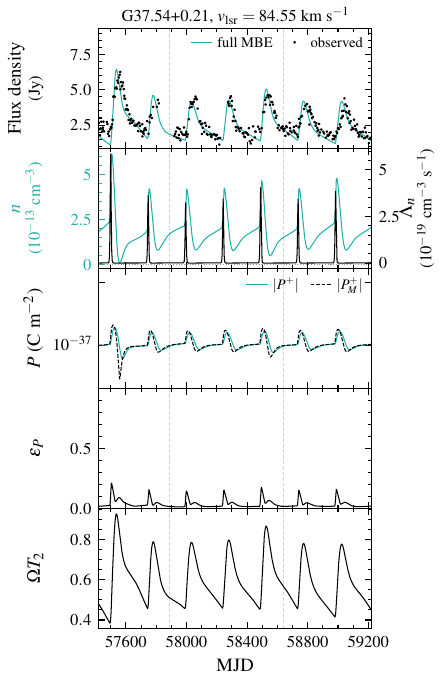}\\[2mm]
\includegraphics[width=0.40\textwidth,height=0.46\textheight,keepaspectratio]{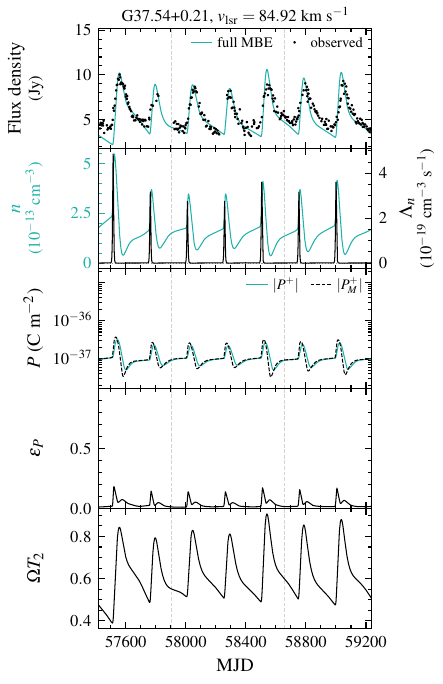}%
\hspace{0.05\textwidth}%
\includegraphics[width=0.40\textwidth,height=0.46\textheight,keepaspectratio]{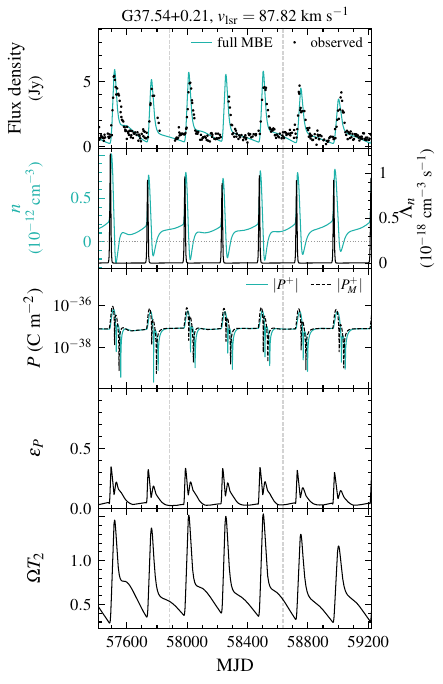}
\caption{Same as Figure~\ref{fig:diag_g12_pair}, for G37.54$+$0.21 at $v_{\rm lsr}=83.77$ (top left), $84.55$ (top right), $84.92$ (bottom left), and $87.82~\mathrm{km\,s^{-1}}$ (bottom right). Vertical dashed lines mark the centres of the two smooth $T_2$ transitions.}
\label{fig:diag_g37_four}
\end{figure}

Figures~\ref{fig:diag_g12_pair}--\ref{fig:diag_g37_four} show the fits and coherence diagnostics for the ten components not illustrated in Figure~\ref{fig:g37_diagnostics}.

\clearpage
\twocolumn


\bibliographystyle{elsarticle-harv}
\bibliography{refs}

\end{document}